\documentclass[12pt,preprint]{aastex}
\usepackage{amsmath, amsthm, amssymb,cases}
\usepackage{natbib}

\ifnum 2<1
\newcommand{\aap}{{Astron. Astrophys.}}
\newcommand{\apj}{{Astrophys. J.}}
\newcommand{\apjl}{{Astrophys. J. Lett.}}

\newcommand{\grl}{{Geophys. Res. Lett.}}
\newcommand{\jgr}{{J. Geophys. Res.}}
\newcommand{\mnras}{{Mon. Not. Roy. Astron. Soc.}}
\newcommand{\nat}{{Nature}}

\newcommand{\solphys}{{Solar Phys.}}

\newcommand{\ssr}{{Space Sci. Rev.}}
\fi

\begin{document}


\title{Predictability of the solar cycle over one cycle}

\author{Jie Jiang\altaffilmark{1,2} Jing-Xiu Wang\altaffilmark{2}, Qi-Rong Jiao\altaffilmark{1}, and Jin-Bin Cao\altaffilmark{1}}
\altaffiltext{1}{School of Space and Environment, Beihang University, Beijing, China}
\altaffiltext{2}{Key Laboratory of Solar Activity, National Astronomical Observatories, Chinese Academy of Sciences, Beijing 100012, China}

\email{jiejiang@buaa.edu.cn}

\begin{abstract}
The prediction of the strength of future solar cycles is of interest because of its practical significance for space weather and as a test of our theoretical understanding of the solar cycle. The Babcock-Leighton mechanism allows predictions by assimilating the observed magnetic field on the surface. But the emergence of sunspot groups has the random properties, which make it impossible to accurately predict the solar cycle and also strongly limit the scope of cycle predictions. Hence we develop the scheme to investigate the predictability of the solar cycle over one cycle. When a cycle has been ongoing for more than 3 years, the sunspot group emergence can be predicted along with its uncertainty during the rest time of the cycle. The method for doing this is to start by generating a set of random realizations which obey the statistical relations of the sunspot emergence. We then use a surface flux transport model to calculate the possible axial dipole moment evolutions. The correlation between the axial dipole moment at cycle minimum and the subsequent cycle strength and other empirical properties of solar cycles are used to predict the possible profiles of the subsequent cycle. We apply this scheme to predict the large-scale field evolution from 2018 to the end of cycle 25, whose maximum strength is expected to lie in the range from 93 to 155 with a probability of 95\%.
\end{abstract}

\keywords{Sun: magnetic fields, Sun: activity}

\section{Introduction}
\label{sec:introduction}

The various proxies of solar activity show a roughly 11-year cycle period with widely varying amplitudes. The cycle amplitude gives a rough idea of the frequency of space weather storms of all types, from radio blackouts to geomagnetic storms to radiation storms. For this reason there is a practical need for solar cycle prediction in our technological society. The longer the prediction extends with higher reliability, the more desirable it is for decision makers. Moreover, the predictions, especially the dynamo-based prediction, can also provide us the opportunity to examine our insight into the physical mechanisms underlying the solar cycle.

Existing attempts to predict the level of solar activity can be broadly divided into two groups: (1) those predicting an ongoing cycle, and (2) those predicting the next or future cycle(s). The first group can be further divided into medium-term predictions (months in advance) and long-term predictions (years in advance). Three methods for short term  prediction of ongoing cycles are the McNish-Lincoln method \citep{McNish1949}, the standard method, and the combined method at Sunspot Index and Long-term Solar Observations (SILOS) \footnote{http://www.sidc.be/silso/forecasts}. For the long-term ongoing cycle predictions, people usually use the curve-fitting function \citep{Waldmeier1935, Hathaway1994, Li1999, Du2011a, Li2017} or use the similarity of solar cycles, e.g., \cite{Li2002}.

For the long-term predictions of the future cycle(s), existing attempts can be broadly divided into empirical methods and methods based on dynamo models. The empirical methods can be further divided into two subgroups. One group is based on extrapolating the past records using a purely mathematical or statistical analysis, which has wide applications in econometrics \footnote{https://robjhyndman.com/hyndsight/fpp-e-book/}. The other subgroup of the empirical methods are the precursor methods, which are based on correlations between certain measured quantities in the declining phase of a cycle and the strength of the next cycle, e.g., \cite{Thompson1993, Svalgaard2005, Cameron2007}. Around the end of cycle 23, two group of people made the dynamo-based solar cycle prediction available for the first time \citep{Choudhuri2007, Jiang2007, Dikpati2006a, Dikpati2006b}. The different observed data assimilated into prediction models and different understandings of the flux transport mechanisms cause markedly distinct predictions for cycle 24 \citep{Karak2014}. For a review about the solar cycle prediction, see \cite{Petrovay2010} and references therein.

Although there have been numerous attempts to predict the solar cycle, the reliability of the different methods is still controversial. The predictions of cycle 24 using the different prediction methods, with a comparison to the true strength of solar cycle 24, are clearly presented in \cite{Pesnell2012} who analysed 75 predictions for cycle 24. The proved most successful method is the precursor method based on the polar field \citep{Schatten1978, Schatten1987, Layden1991, Schatten1996, Svalgaard2005} and the geomagnetic activity \citep{Ohl1979, Feynman1982}, which is a good proxy for the Sun's polar field during cycle minima \citep{Wang2009}.

The correlation of the polar field at cycle minimum on the subsequent cycle strength is the natural result of the Babcock-Leighton (BL)  dynamo \citep{Babcock1961, Leighton1969}. Recently \cite{Cameron2015} demonstrated that the net toroidal magnetic flux generated by differential rotation within a hemisphere of the convection zone is determined by the emerged magnetic flux at the solar surface. The demonstration and the predictive skill of the polar field indicate that the solar dynamo is of the BL type. In the BL dynamo framework, the poloidal field regeneration, half of the dynamo loop, is accessible to direct observation. Furthermore, the unavoidable time taken to  wind up  the surface poloidal, means that surface fields observed today will be the source of the sunspot-forming toroidal magnetic component in the future. This justifies the method to use the polar field at cycle minimum to produce a cycle prediction. But can we predict the polar field evolution before cycle minimum so that the prediction scope can be extended?

The polar field evolution can be understood in terms of the Surface Flux Transport (SFT) model \citep{Wang1989, Mackay2012,Jiang2014a}. The tilt angle of the sunspot group with respect to the E-W direction, leads to each active region contributing to the global axial dipole moment of the Sun, and is an essential factor in the polar field generation. The tilt angle has a systematic component, which corresponds to the Joy's law, and a larger random component, i.e., the tilt angle scatter. \cite{Jiang2014b} found the observed scatter of the tilt angles causes a variation of 30\%-40\% in the resulting polar field around activity minima. By including detailed information about the individual tilt angles and magnetic polarities of the bipolar magnetic regions (BMRs) that emerged during cycle 23, the weak polar field around the end of cycle 23 is well reproduced by \cite{Jiang2015}. Recently series of BL dynamo models successfully reproduce the variability of the solar cycle by introducing the random BMR tilts \citep{Cameron2017,Olemskoy2013b, Jiang2013,Lemerle2017,Nagy2017,Karak2017}. The randomness of the sunspot emergence -- part of the toroidal-to-poloidal part of the dynamo loop -- makes the dynamo a stochastic process. Even if the poloidal-to-toroidal part of the dynamo loop was fully deterministic, the intrinsic random features of flux emergence limits the scope of the solar cycle prediction. Furthermore, the randomness of the sunspot emergence due to the complex flux emergence dynamics \citep{Weber2011,Weber2013} also causes the uncertainty even for the prediction of an ongoing cycle. The randomness raises the question of whether the long-term predictions with reasonable levels of uncertainty are possible. A second question is the extent to which we can predict the time-dependence of the activity of a solar cycle beyond only predicting the cycle amplitude.

\cite{Jiang2011} show the dependence of the statistical properties of sunspot emergence on the cycle phase and strength based on an empirical analysis of the historical sunspot number data \citep[see also][]{Munoz-Jaramillo2015}. Random realizations of sunspot group emergences during a cycle can be reconstructed for a given maximum sunspot number based on the statistical properties. The SFT model can be used to produce the large-scale field evolution over the Sun's surface. \cite{Cameron2016b} and \cite{Jiang2017} have applied the above scheme in a Monte-Carlo approach to predict the range of the polar field around the end of cycle 24 about 3-4 years before the minimum, including error bars. Similar approaches are also presented in \cite{Hathaway2016} and \cite{Iijima2017}. In this paper, we generalize and update the previous method to investigate the predictability of the solar cycle over one cycle at different phases of a cycle. The predictions include the time evolution of the monthly sunspot number of an ongoing cycle and the time evolution of the smoothed sunspot number over one cycle, all with estimated uncertainties. The investigation of the predictability beginning from different starting points distinguishes our scheme from the existing predictions of the solar cycle.

The paper is organized as follows. In Section 2, we study the properties of the solar cycle profiles, which will be adopted by Section 3 for the prediction of sunspot emergence for an ongoing cycle. The SFT model which will be used to describe and to predict the large-scale field evolution over the solar surface is presented in Section 4. The correlation between the dipole moment at cycle minimum and the subsequent cycle strength based on a homogeneous axial dipole moment dataset is given in Section 5. In Section 6, we present the results about the predictability of the subsequent cycle and its application to cycle 25. Our summary and discussion are given in Section 7.

\section{Properties of solar cycle profiles}
\label{sec:CycleShape}
At present there are different time series of the sunspot number, which differ substantially before the cycle 12 \citep{Clette2014}. Hence in this study we only investigate the properties of solar cycle profiles during cycles 12-24, and have used the Sunspot Number Version 2.0 \footnote{http://www.sidc.be/silso/datafiles}. The data are of  the monthly mean total sunspot number ($R_{\textrm{mn}}$) and the corresponding 13-month smoothed sunspot numbers ($R_{\textrm{sm}}$) . The timing of each cycle minimum used throughout the paper is taken from the NGDC \footnote{https://www.ngdc.noaa.gov/stp/space-weather/solar-data/solar-indices/sunspot-numbers/cycle-data/}. We first analyse the properties of the smoothed sunspot numbers. The monthly sunspot number will be analyzed at the end of this section.

\begin{figure}[!htp]
 \centering
\includegraphics[scale=0.4]{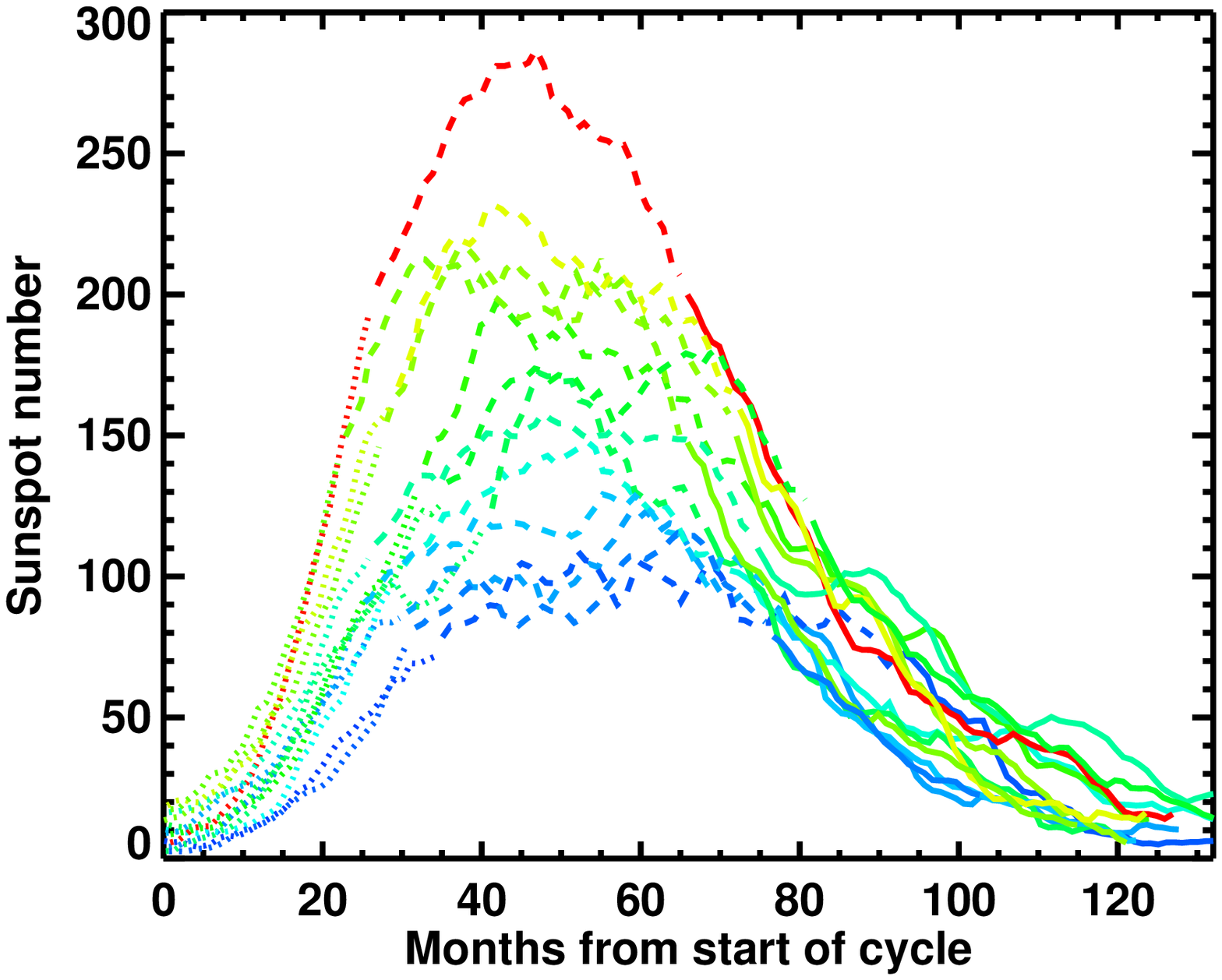}
\caption{Sunspot number as a function of time from minimum activity for cycles 12-24 in different colors. The rising phase, the maximum phase and the declining phase are distinguished in dotted, dashed and solid curves, respectively.}
\label{fig:AllCycles}
\end{figure}

\begin{figure}[!htp]
 \centering
\includegraphics[scale=0.5]{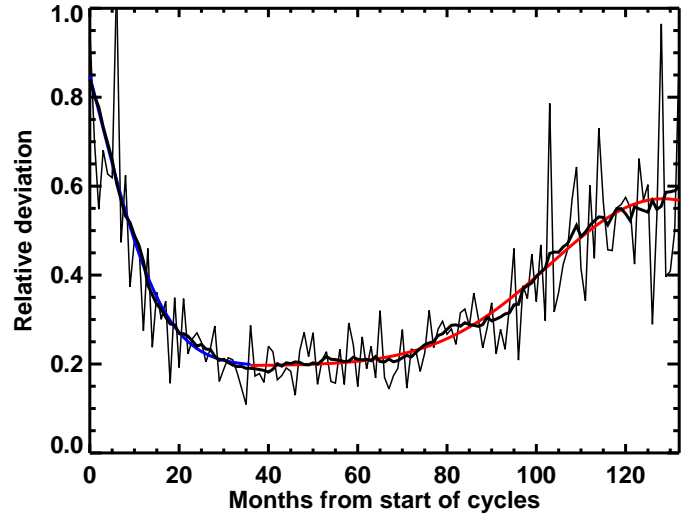}
\caption{Relative deviations of the monthly sunspot number $R_{\textrm{mn}}$ from the smoothed sunspot number $R_{\textrm{sm}}$. The thin black curve is the $s_{\textrm{mn2sm}}$ values. The thick black curve is the 13 month average of the value. The red and blue curves correspond to the curve fits to the first 3 yrs and later 8 years of solar cycles based on Eq.(\ref{eq:s2}).}
\label{fig:SNsm2monthStd}
\end{figure}

Figure \ref{fig:AllCycles} shows the observed smoothed sunspot number as a function of months from start of a cycle for cycles 12-24. We define the maximum phase of a cycle as the time period when the sunspot number surpasses 70\% of maximum sunspot number of the cycle. The ascending phases, the maximum phases, and the declining phases are denoted in the dotted, the dashed, and the solid curves in Figure \ref{fig:AllCycles}, respectively. There are two distinctive features about the shape of the solar cycle. The first is about the rising phase, which obeys the Waldmeier effect \citep{Waldmeier1955}. It is that stronger cycles tend to show a faster rise of activity levels during their ascending phase than weaker cycles \citep{Lantos2000, Cameron2008}. The second is that once the solar cycle begins to decline, all cycles decline in a similar way \citep{Ivanov2014,Cameron2016}. A large scatter of the individual cycle about the means over all the cycles is also shown. During the declining phase, the profile of solar cycle shows noisier short-term variation than the early time period. These properties can be understood under the framework of the BL dynamo. A weaker polar field at the beginning of a cycle generates less toroidal field to form the sunspot groups than it would if the polar field was strong. This provides an explanation to the Waldmeier effect and also is demonstrated by \cite{Karak2011} using a BL dynamo model. \cite{Cameron2016} interpret the similar decline phases in terms of oppositely directed toroidal flux bands in each hemisphere that diffuse and cancel across the equator when the distance of the center of the activity belts from the equator becomes about equal to their width. Stronger cycles show wider activity belts and thus start to decline earlier than weaker cycles. \cite{Nagy2017} and \cite{Kitchatinov2018} showed that peculiar BMRs with large tilt angles emerging during the rising phase of a cycle have large effects on the amplitude of the descending phase. Hence, the stochastic mechanism due to random emergence of the peculiar BMRs during the rising phase causes the later  phases of the cycle to be noisy, as seen in Figure \ref{fig:SNsm2monthStd}.

Several of these properties have been exploited in the development of a parameterization of the sunspot number during a cycle in terms of two simple paramters, the starting time ($t_0$) and amplitude ($a$) \citep[as given in][hereafter HWR94]{Hathaway1994}. The determination of the two parameters can be made a few years after the start of a cycle. The function captures the essence of the Waldmeier effect. The rising phase is presented in the form of $t^3$, where $t$ is the time in months from the start of a cycle. We use the function, i.e. Equation (1) of HWR94 to fit the cycles 12-24 individually. It is in the form of
\begin{equation}\label{eq:cyFit}
f(t)=\frac{a(t-t_0)^3}{\exp[(t-t_0)^2/b^2]-c},
\end{equation}
where $b(a)=27.12+25.15/(a\times10^3)^{1/4}$ and $c=0.71$. The solar cycle overlap, i.e., new cycle spots appearing at high latitudes while old cycles still seen at low latitudes, is about 3 years \citep{Wilson1988,Harvey1992}. We set weights to the first 1.5 year of each cycle by a error function in the form of $\frac{1}{2}[1+\textrm{erf}(2\frac{t-t_{\textrm{ovp}}}{t_{\textrm{ovp}}})]$, where $t_{\textrm{ovp}}$=0.75yr to decrease the effect of the data during the first 1.5 year on the cycle fit.

\begin{figure}[!htp]
 \centering
\includegraphics[scale=0.5,angle=90]{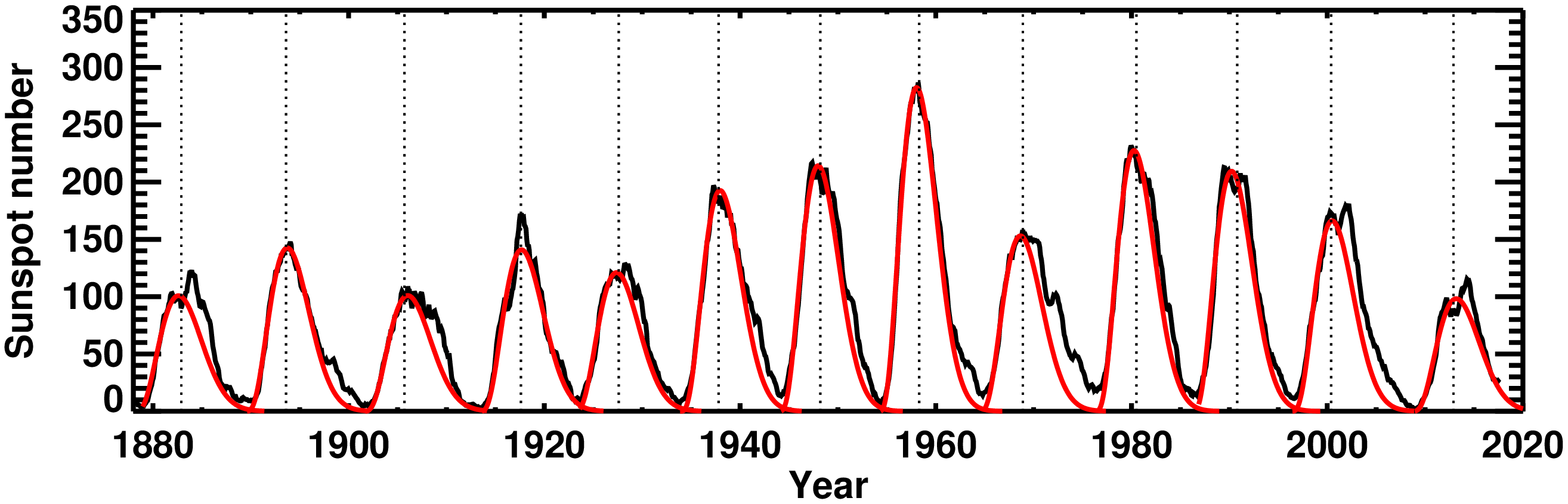}
\includegraphics[scale=0.5,angle=90]{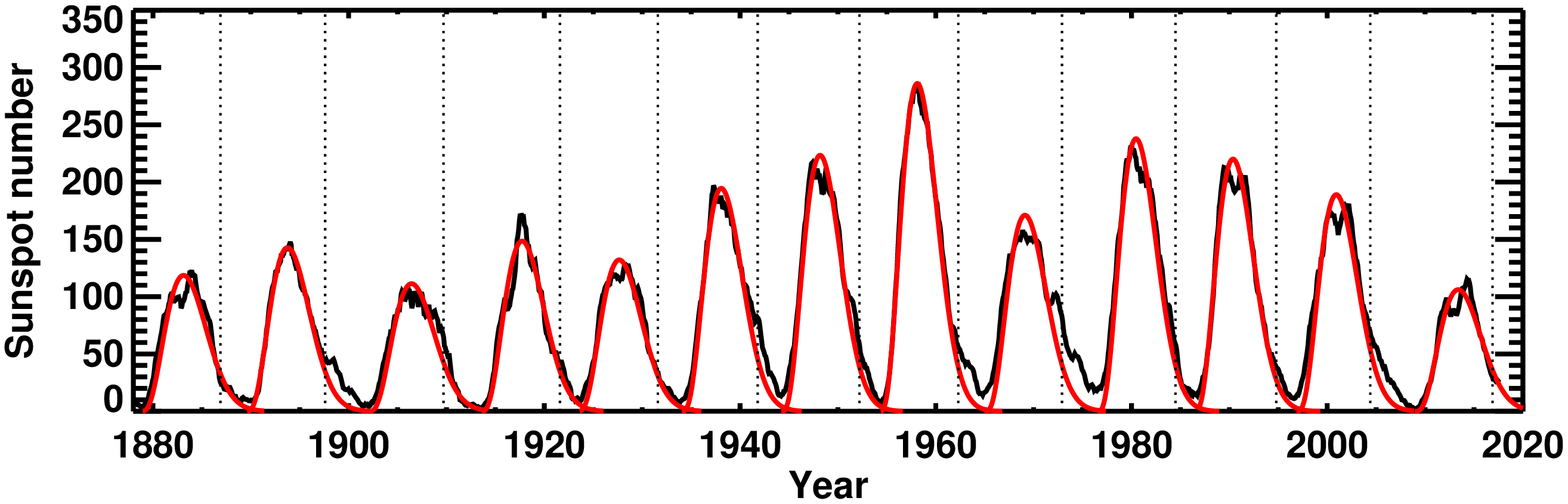}
\caption{Examples to use Eq.(\ref{eq:cyFit}) to fit the shape of solar cycles from 1878 to the present at different phases of solar cycles denoted by the dashed vertical lines. Observations are in black curves and functional fits are in red curves. Upper panel: fits at 4 years into each cycle; Lower panel: fits at 8 years into each cycle.}
\label{fig:CyFit}
\end{figure}

\begin{figure}[!htp]
 \centering
\includegraphics[scale=0.5]{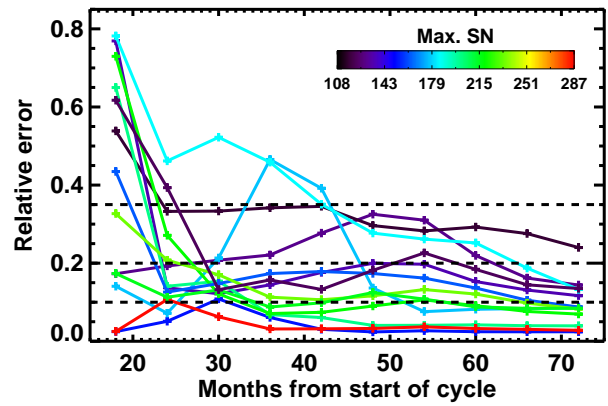}
\caption{Relative error $s_{f2o}$ of the functional fits by Eq.(\ref{eq:cyFit}) comparing with the observations during the maximum phases of solar cycles in different colors when the fits are done at different times (from 18 months to 72 months into each cycle with 6 months interval) after the start of the cycle. The stronger cycles in green to red colors prefer to have smaller values and the weaker cycles tend to have large values.}
\label{fig:cycles_fits_MaxPhases}
\end{figure}

Figure \ref{fig:CyFit} shows examples of the cycle fit overplotted with the observed sunspot number evolution from 1878 to the present, based on the fit either 4 years (upper panel), or 8 years (lower panel), after the start of the cycle. Since the function captures the essence of the Waldmeier effect, the fit and smoothed sunspot number are consistent with each other during the ascending phase. We measure the relative error $s_{\textrm{f2o}}$ between the fit and the observation during the maximum phases for different cycles when the fit is done at different timings $t$ of cycles. Here $s_{\textrm{f2o}}$ is defined as
\begin{equation}\label{eq:relativeError}
s_{\textrm{f2o}}=\sqrt{\frac{\sum_{i=1}^{N}(\frac{R_{\textrm{sm}}(t_i)-f(t_i)}{f(t_i)})^2}{N}},
\end{equation}
where $N$ is the number of months of the maximum phase for different cycles. The results are shown in Figure \ref{fig:cycles_fits_MaxPhases}. The fits to cycles 12-24 are done from 1.5 years to 6 years into cycles with 0.5 year interval. The strength of a cycle defined by the maximum sunspot number is indicated by a given color shown by the color bar. The stronger cycles shown in colors between the green to the red tend to have smaller relative errors. They are usually less than 20\% even the fits start from 2 years onwards. The weaker cycles shown in colors between the black to the cyan tend to match with the fits more poorly. From 3 years into a cycle, $s_{\textrm{f2o}}$ is always within 50\%. This leads us to only make ongoing cycle prediction \textbf{from} at least 3 years into a cycle throughout the rest of this paper.

During the later phases, the fitting curves in Figure \ref{fig:CyFit} shows larger deviations from the observations. BMRs during the descending phase tend to locate at low latitudes. \cite{Jiang2014b} demonstrated that the BMRs with the same initial axial dipole moment at lower latitudes have larger contributions to the polar field. Hence the BMRs during the descending phase play important roles in the polar field evolution. This motivates us to suggest an improvement in order  to improve the predictions. We use two methods to measure the differences $\triangle f(t)$ between the observation $R_{\textrm{sm}}(t)$ and the fitted function $f(t)$. The first is \begin{equation}
R_{\textrm{sm}}(t)=f(t)+\triangle f(t),
\end{equation}
where
\begin{equation}\label{eq:fit1}
\triangle f(t)=\overline{\triangle f(t)}+\triangle f'(t),
\end{equation}
where $\overline{\triangle f(t)}$ will be added to function (\ref{eq:cyFit}) to correct \textbf{the systematic deviation from} the cycle fits especially in the descending phase. Meantime, we estimate the fluctuation of the shape of cycles by measuring the standard deviation ($\sigma_{\triangle f'(t)}$) of fits including the correction using $\overline{\triangle f(t)}$ from observations. The $\overline{\triangle f(t)}$ and $\sigma_{\triangle f'(t)}$ are calculated by
\begin{equation}\label{eq:rstd1}
\overline{\triangle f(t)}=\frac{\sum_{i=1}^{N} \left( R_{\textrm{sm}}(t)_i-f(t)_{i}\right)}{N}, ~and
\end{equation}
\begin{equation}\label{eq:rstd1}
\sigma_{\triangle f'(t)}=\sqrt{ \frac{\sum_{i=1}^{N} \left( R_{\textrm{sm}}(t)_i-f(t)_{i}-\overline{\triangle f(t)}_{i} \right)^2 }{N} },
\end{equation}
where the symbol $i$ denotes different cycles 12-24 and $N$ is equal to 13. The random component $\triangle f'(t)$ will be added to predict the sunspot number evolution as a set of random realizations with zero average and the standard deviation of $\sigma_{\triangle f'(t)}$.

The other method to correct the fit is
\begin{equation}R_{\textrm{sm}}(t)=f(t)+\triangle r(t)f(t),
\end{equation}
where
\begin{equation}\label{eq:fit1}
\triangle r(t)=\overline{\triangle r(t)}+\triangle r'(t),
\end{equation}

\begin{equation}\label{eq:fit1}
\overline{\triangle r(t)}=\frac{\sum_{i=1}^{N} \left( \frac{R_{\textrm{sm}}(t)_{i}-f(t)_{i}}{f(t)_{i}} \right)}{N},
\end{equation}
and
\begin{equation}\label{eq:rstd2}
\sigma_{\triangle r'(t)}=\sqrt{ \frac{\sum_{i=1}^{N} \left( \frac{R_{\textrm{sm}}(t)_{i}-f(t)_{i}-\overline{\triangle r(t)}_{i}f(t)_{i}}{f(t)_{i}+\overline{\triangle r(t)}_{i}f(t)_{i}} \right)^2 }{N} },
\end{equation}
where $i$-values are the same as that in Eq.(\ref{eq:rstd1}), which are the 13 cycles from cycle 12 to cycle 24. The random component $\triangle r'(t)$ will be added to predict the sunspot number evolution as a set of random realizations with zero average and the standard deviation of $\sigma_{\triangle r'(t)}$.

\begin{figure}
 \centering
  \includegraphics[width=0.45\textwidth]{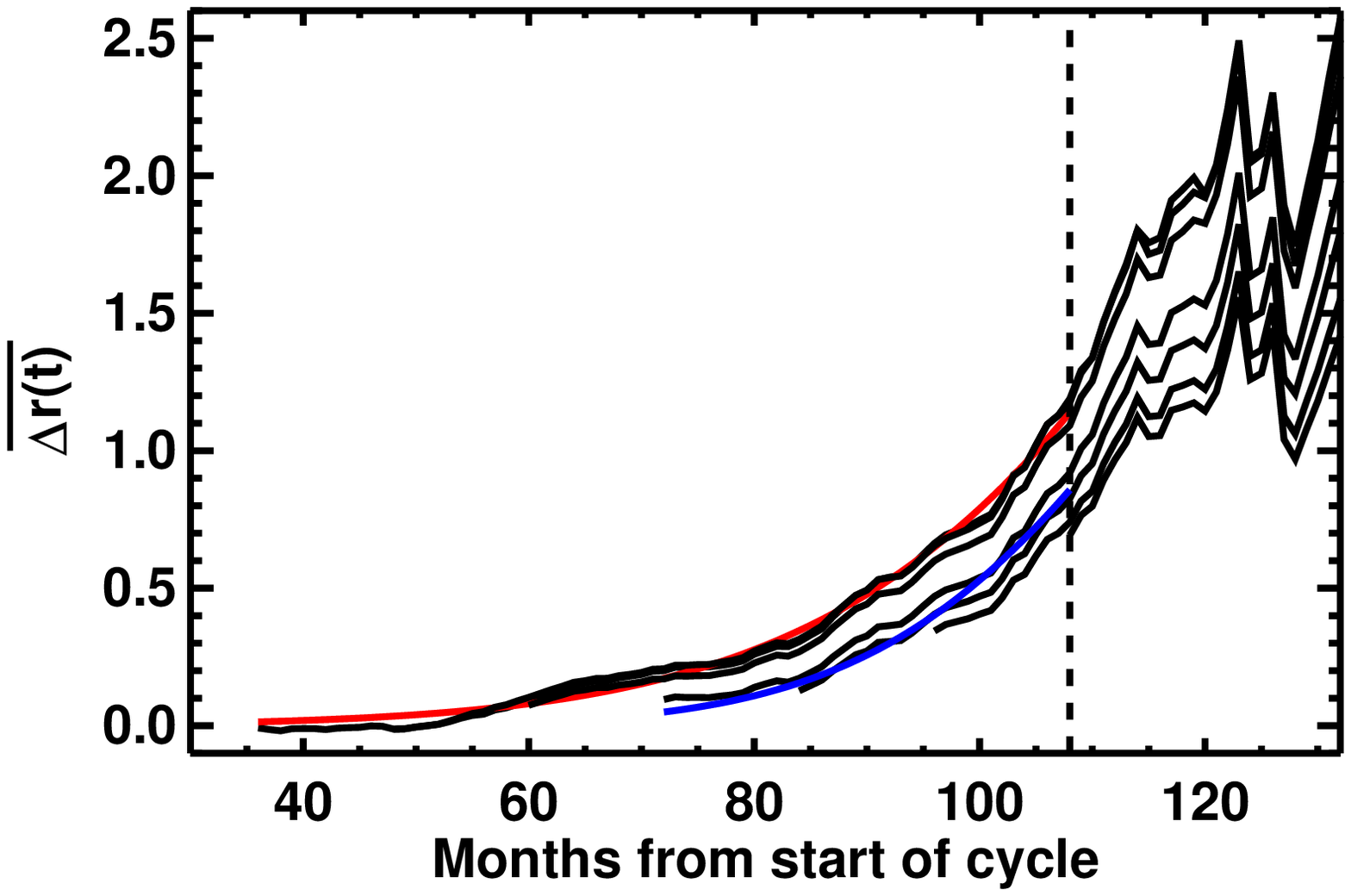}
  \includegraphics[width=0.45\textwidth]{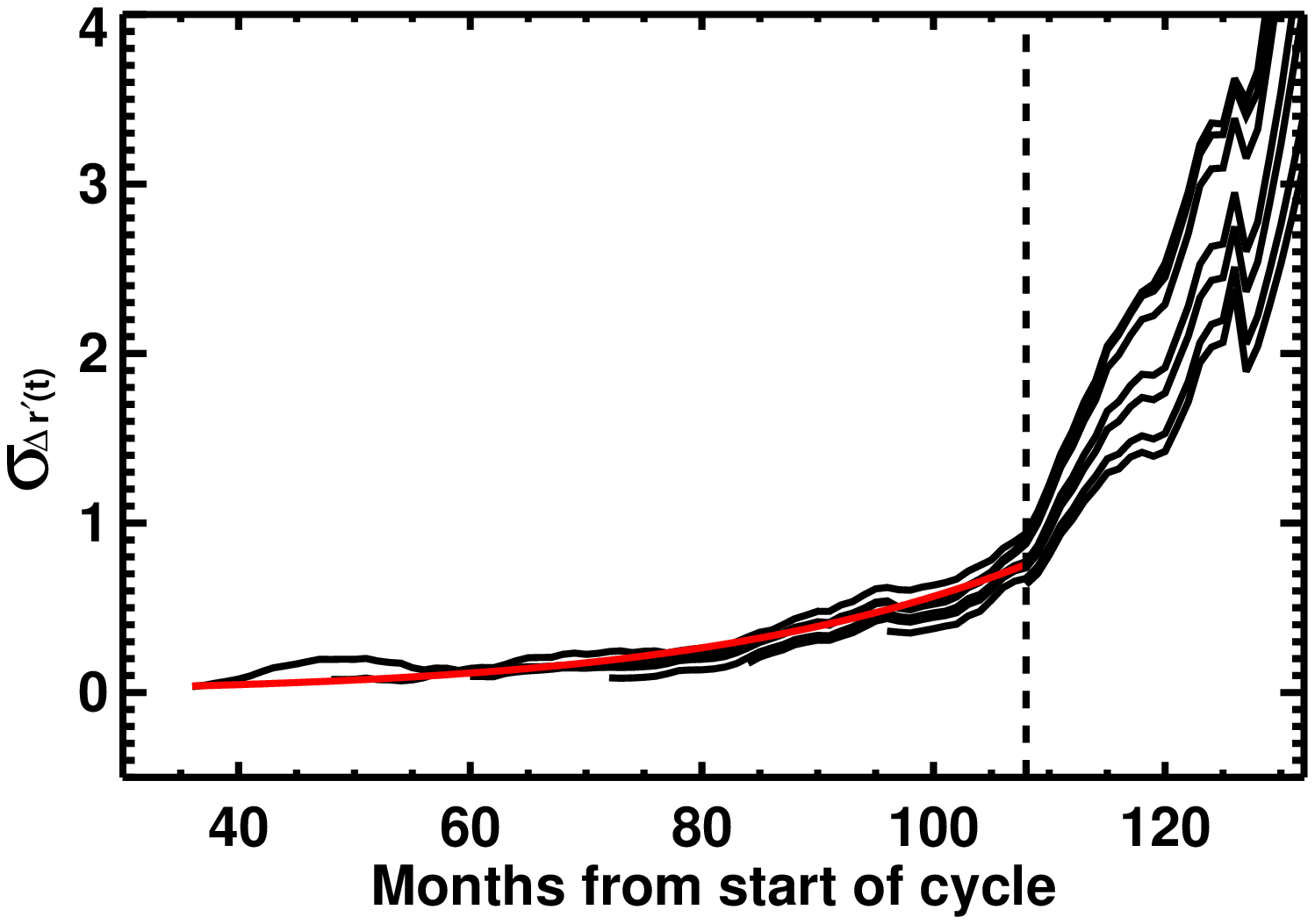}
  \includegraphics[width=0.45\textwidth]{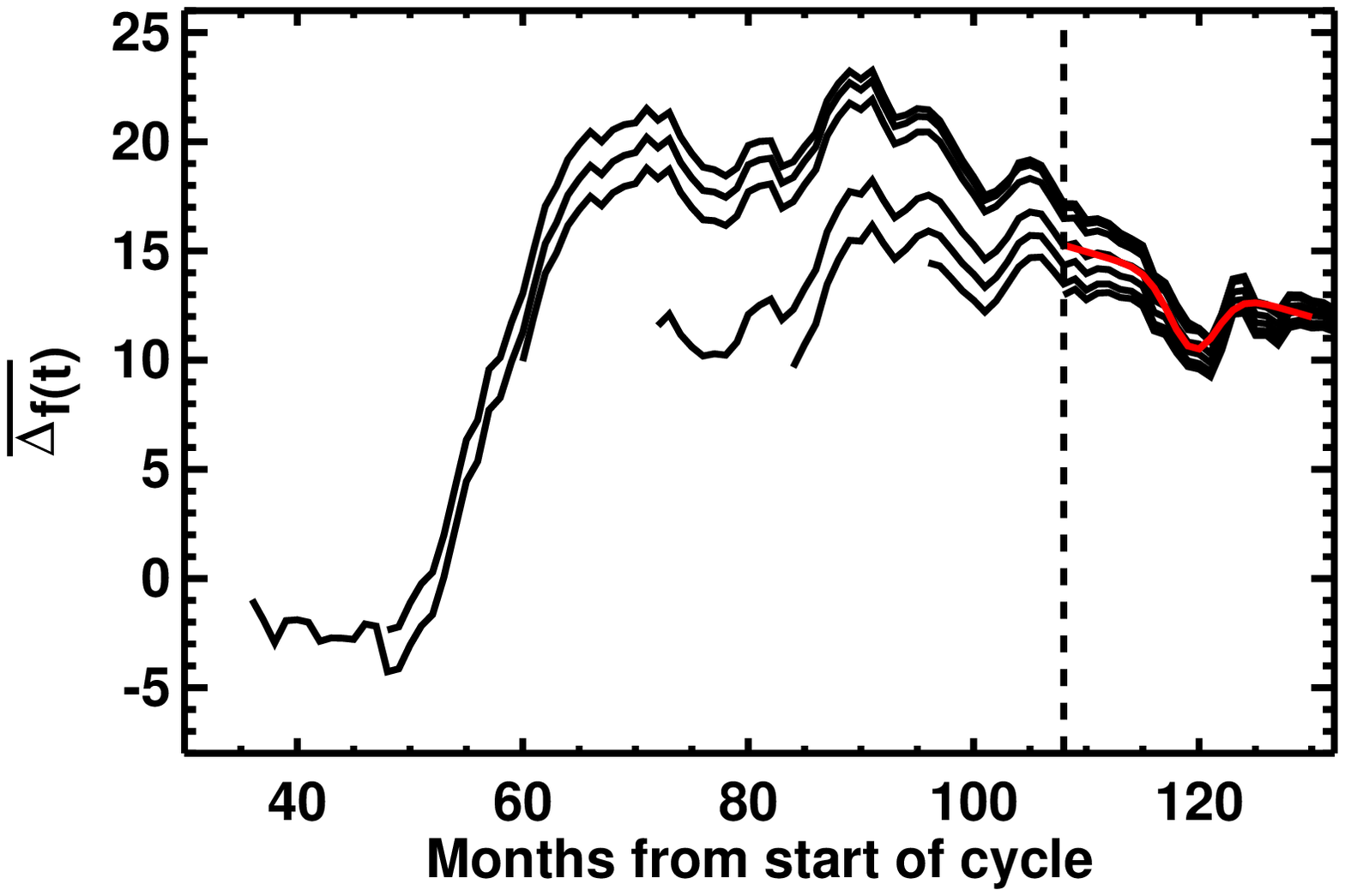}
  \includegraphics[width=0.45\textwidth]{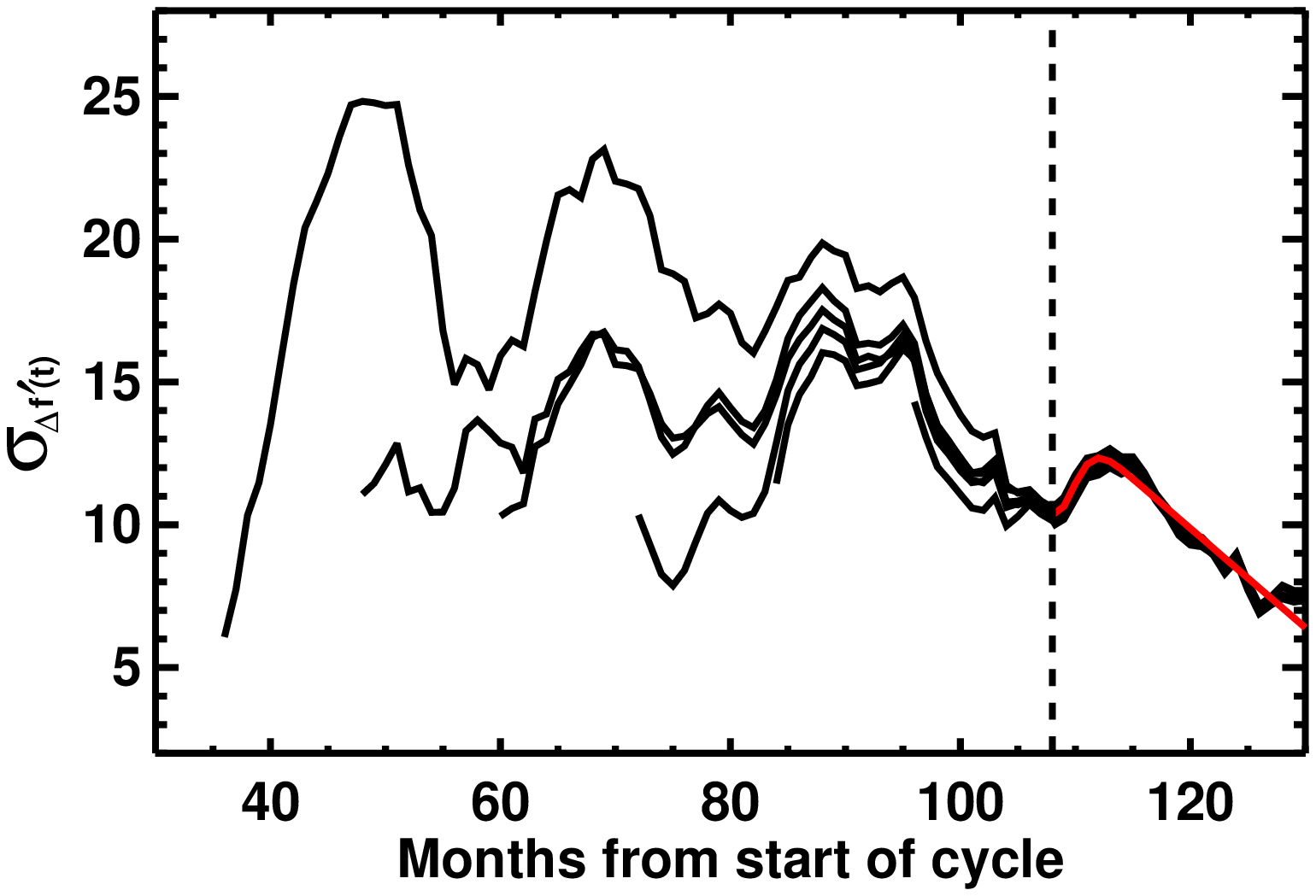}
  \caption{Time dependence of $\overline{\triangle r(t)}$ (upper left), $\sigma_{\triangle r'(t)}$ (upper right), $\overline{\triangle f(t)}$ (lower left), and $\sigma_{\triangle f'(t)}$ (lower right) when the fits by Eq.(\ref{eq:cyFit}) are done at different times after the start of the cycle. Black curves in each panel correspond to values when fits are done during 36 months to 108 months into solar cycles with 12-month intervals. The color curves correspond to the curve fits to the averaged values during different time periods. The vertical lines correspond to the timing of 108 months into solar cycles.}
\label{fig:measureCyFit}
\end{figure}

Figure \ref{fig:measureCyFit} shows the time evolution of $\overline{\triangle f(t)}$, $\sigma_{\triangle f'(t)}$, $\overline{\triangle r(t)}$, and $\sigma_{\triangle r'(t)}$ when the fits are done from 3 years or more into a cycle. The curves in each panel correspond to the different timings (from the 3rd years to the 9th years into cycles with one year cadence) to do the cycle fits. We see that $\overline{\triangle r(t)}$ and $\sigma_{\triangle r'(t)}$ increase for the later phase of cycles. When the fits are done at the 3rd year, the 4th year and the 5th year, $\overline{\triangle r(t)}$ and $\sigma_{\triangle r'(t)}$ have the almost same profiles. If the fits are done at the 6th year to the 8th year, there is a significant increase of the match between the observed values and the fitting function. And the $\triangle r(t)$-values decrease accordingly. Close to the minimum time period, i.e., from the 9th year into a cycle, there are large deviations among different fits. The small sunspot number during the decline phase contributes to the large variations of the $\triangle r(t)$-value. On the contrast, $\overline{\triangle f(t)}$ and $\sigma_{\triangle f'(t)}$ during this time period weakly depend on the fits which are done at different phases of a cycle.

The fitting functions for the time evolution of $\overline{\triangle r(t)}$ and $\sigma_{\triangle r'}(t)$ are listed in the following.
\begin{equation}\label{eq:r1a}
\overline{\triangle r(t)}=8.68\exp(-z_1^2/2), t < 72~\mathrm{months}
\end{equation}
\begin{equation}\label{eq:r1b}
\overline{\triangle r(t)}=2.64\exp(-z_2^2/2), 72~\mathrm{months} \leq t\leq 108~\mathrm{months},
\end{equation}
where $z_1=\frac{t-200.44}{45.86}$ and $z_2=\frac{t-149.0}{27.31}$. That Eq.(\ref{eq:r1a}) or Eq.(\ref{eq:r1b}) is used depends on when the prediction is done. When it is less than 72 months into a cycle, we take Eq.(\ref{eq:r1a}). When it is more than 72 months and less than 108 months into a cycle, we take Eq.(\ref{eq:r1b}). The corresponding $\sigma_{\triangle r'(t)}$ has a weak dependence on the fitting time and satisfies the following form:
\begin{equation}\label{eq:sigma_r}
\sigma_{\triangle r'(t)}=30.92\exp(-z_3^2/2), t\leq 108~\mathrm{months}\\,
\end{equation}
where $z_3=\frac{t-318.73}{77.35}$. During the later phases, the $\overline{\triangle f(t)}$ obeys the following profiles,
\begin{equation}\label{eq:dft}
\overline{\triangle f(t)}=-2.41\exp(-z_4^2/2)+36.8-0.2t,  t\geq 108~\mathrm{months}\\,
\end{equation}
where $z_4=\frac{t-119.3}{2.0}$.
The corresponding $\sigma_{\triangle f'(t)}$ is
\begin{equation}\label{eq:d_sigma_ft}
\sigma_{\triangle f'(t)}=-3.06\exp(-z_5^2/2)+47.23-0.314t,  t\geq 108~\mathrm{months}\\.
\end{equation}
where $z_5=\frac{t-108.16}{1.455}$. The two parameters $\overline{\triangle f(t)}$ and $\sigma_{\triangle f'(t)}$ are shown in the red curves in the lower left and the lower right panels, respectively.

Now we analyze the monthly sunspot number $R_{\textrm{mn}}$. The deviation of the $R_{\textrm{mn}}$ from the smoothed sunspot number $R_{\textrm{sm}}$ is regarded as a random component. We measure the relative deviations $s_{\textrm{mn2sm}}$ of $R_{\textrm{mn}}$ from $R_{\textrm{sm}}$. It is defined as
\begin{equation}\label{eq:relativeError2}
s_{\textrm{mn2sm}}=\sqrt{\frac{\sum_{i=1}^{N}(\frac{R_{\textrm{mn}}(t)_i-R_{\textrm{sm}}(t)_i}{R_{\textrm{sm}}(t)_i})^2}{N}},
\end{equation}
where $i$ is the 13 cycles from cycle 12 to 24. The result is shown in Figure \ref{fig:SNsm2monthStd}. We see during the maximum phase the relative deviations $s_{\textrm{mn2sm}}$ keep at a small value of about 0.2. Then the value increases with time during the decline phase, which means noisier later phases of solar cycle than earlier phases. The reason has been given in the second paragraph of this section. During the first 2 years, the large $s_{\textrm{mn2sm}}$ is due to the cycle overlap. The fitting functions are
\begin{equation}\label{eq:s2}
s_{\textrm{mn2sm}}=\begin{cases}
0.936\exp(-p_1^2/2)+0.196, t < 36~\mathrm{months},\\
0.375\exp(-p_2^2/2)+0.196, t \geq 36~\mathrm{months}, \\
\end{cases}
\end{equation}
where $p_1=\frac{t+12.21}{14.346}$ and $p_2=\frac{t-128.4}{25.357}$.

\section{Prediction of sunspot emergence of an ongoing cycle}
\label{sec:PrdcEmergence}
\subsection{Prediction of the sunspot number evolution}
With the features of the solar cycle shape listed in Section \ref{sec:CycleShape}, we may predict the time evolution of the monthly sunspot number $F_{\textrm{mn}}$(t) for the remainder of a cycle when it is some months, denoted as $n$, entering into the cycle. We separate the sunspot number into the \textit{systematic component} and the \textit{random component}. We first use the method described in Section \ref{sec:CycleShape} to fit the observed smoothed sunspot number to derive the function $f$(t). Then we get the predicted systematic part of smoothed sunspot number evolution $\overline{F_{\textrm{sm}}}(t)$=$f(t)+\overline{\triangle f(t)}$ or $f(t)+\overline{\triangle r(t)}f(t)$ for the different values of $n$ according to Eqs.(\ref{eq:r1a}), (\ref{eq:r1b}), and (\ref{eq:dft}). The random deviations from the systematic component measured by Eqs.(\ref{eq:sigma_r}) and (\ref{eq:d_sigma_ft}) are determined by using Monte-Carlo simulations. The random realization of the sunspot emergence is denoted as $F_{\textrm{sm}}(t)$, which is equal to $\overline{F_{\textrm{sm}}}(t)+\triangle f'(t)$ or $\overline{F_{\textrm{sm}}}(t)+\triangle r'(t)f(t)$, where $\triangle f'(t)$ and $\triangle r'(t)$ have zero averages, and their standard deviations satisfy Eqs.(\ref{eq:sigma_r}) and (\ref{eq:d_sigma_ft}). The monthly sunspot number $F_{\textrm{mn}}(t)$ is ${F_{\textrm{sm}}}(t)+r_{\textrm{sm}}(t){F_{\textrm{sm}}}(t)$, where the standard deviation of $r_{\textrm{sm}}$ satisfies Eq.(\ref{eq:s2}) and the average of $r_{\textrm{sm}}$ is zero.

We take two timings, 4 years and 8 years into cycles, as examples to compare the differences between the predictions of ongoing cycles $\overline{F_{\textrm{sm}}}(t)$ and the observations $R_{\textrm{sm}}(t)$. The difference is measured by the goodness-of-fit, which is given by
\begin{equation}\label{eq:GoodnessOfFit}
\chi=\sqrt{ \frac{\sum_{i=1}^{N} \left( \frac{\overline{F_{\textrm{sm}}}(t_i)-R_{\textrm{sm}}(t_i)}{\sigma(t_i)} \right)^2 }{N} },
\end{equation}
where $\sigma(t_i)$ corresponds to Eq.(\ref{eq:rstd1}) and $N$ corresponds to the numbers of the months from the timing of the prediction to the end of the cycle. When $\chi$ is equal to 1.0, it indicates that the fitting function passes within one standard deviation of the observed data points. Table \ref{tab:GoodnessOfFit} shows $\chi$ values of the predicted results of the sunspot number evolution during cycles 12-24 at 4 years and 8 years into cycles. The $\chi$ values are also calculated for the prediction just based on HWR94 method, i.e., fits using Eq.(\ref{eq:cyFit}). We see that $\chi$ values are always much smaller than that of HRW94, especially when the predictions are at the earlier phase, i.e., 4 years into a cycle. This demonstrates the improvement of the predictive skill by the current strategy comparing with HWR94.

Another advantage of the method is to quantify the uncertainty of the predictions, which corresponds to the predictability of the sunspot number evolution.  Tables \ref{tab:phases_rstd_sm} and  \ref{tab:phases_rstd} give the evaluation of the predictive skill of the smoothed sunspot number ${F_{\textrm{sm}}}(t)$ and the monthly sunspot number $F_{\textrm{mn}}(t)$, respectively. The values show the percentages of the observed sunspot numbers, which are out of the 1$\sigma$, 2$\sigma$ and 3$\sigma$ of the predicted variations. For some cycles, e.g., cycle 12 and cycle 20, the percentages are larger than 32\% and 4.6\% for 1$\sigma$ and 2$\sigma$ fluctuations respectively. But all the observed sunspot numbers are within 3$\sigma$ range of the prediction. The predictive skill for monthly sunspot number is worse than that of the smoothed sunspot number.

\begin{figure}
 \centering
  \includegraphics[width=0.32\textwidth]{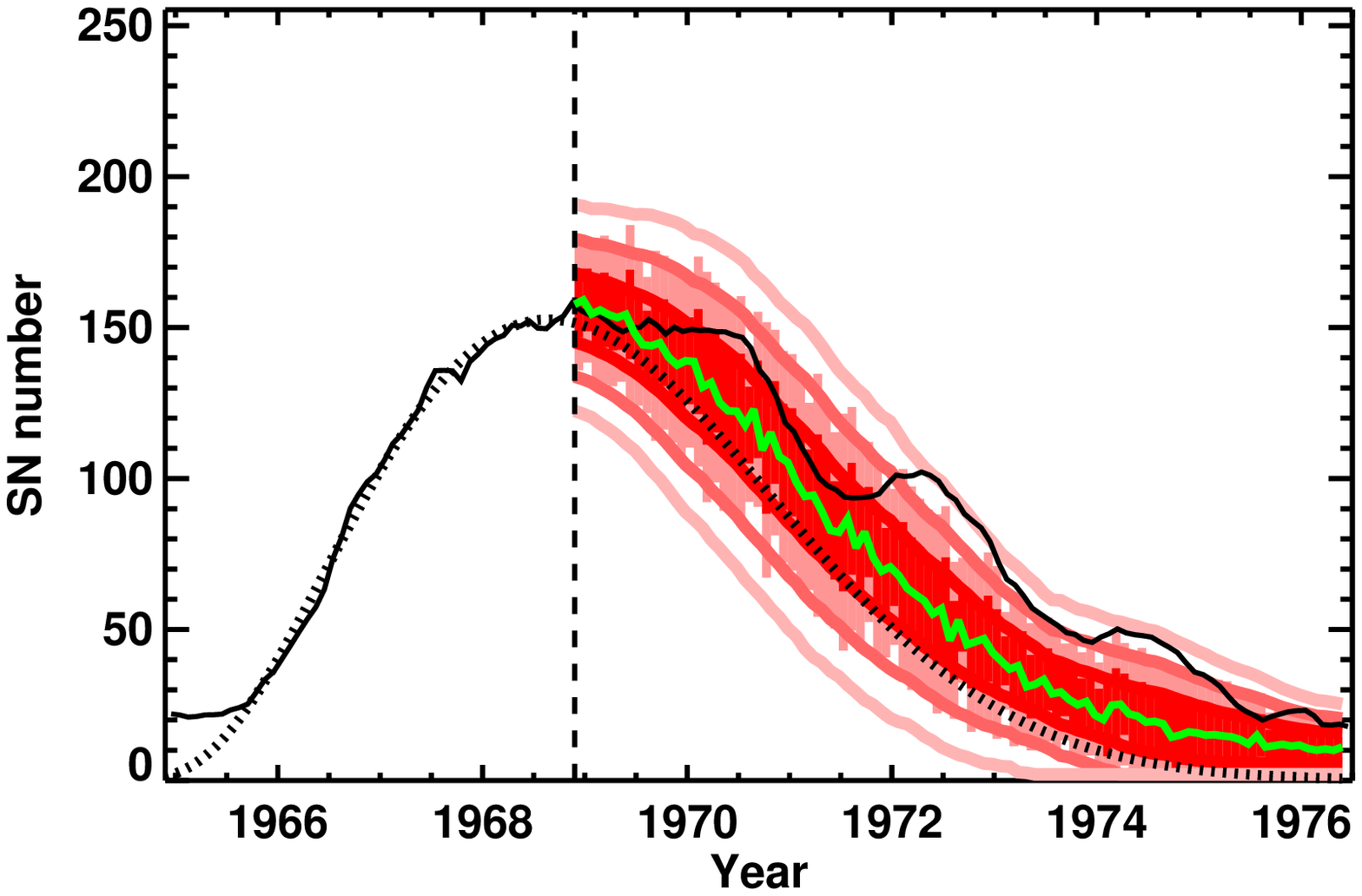}
  \includegraphics[width=0.32\textwidth]{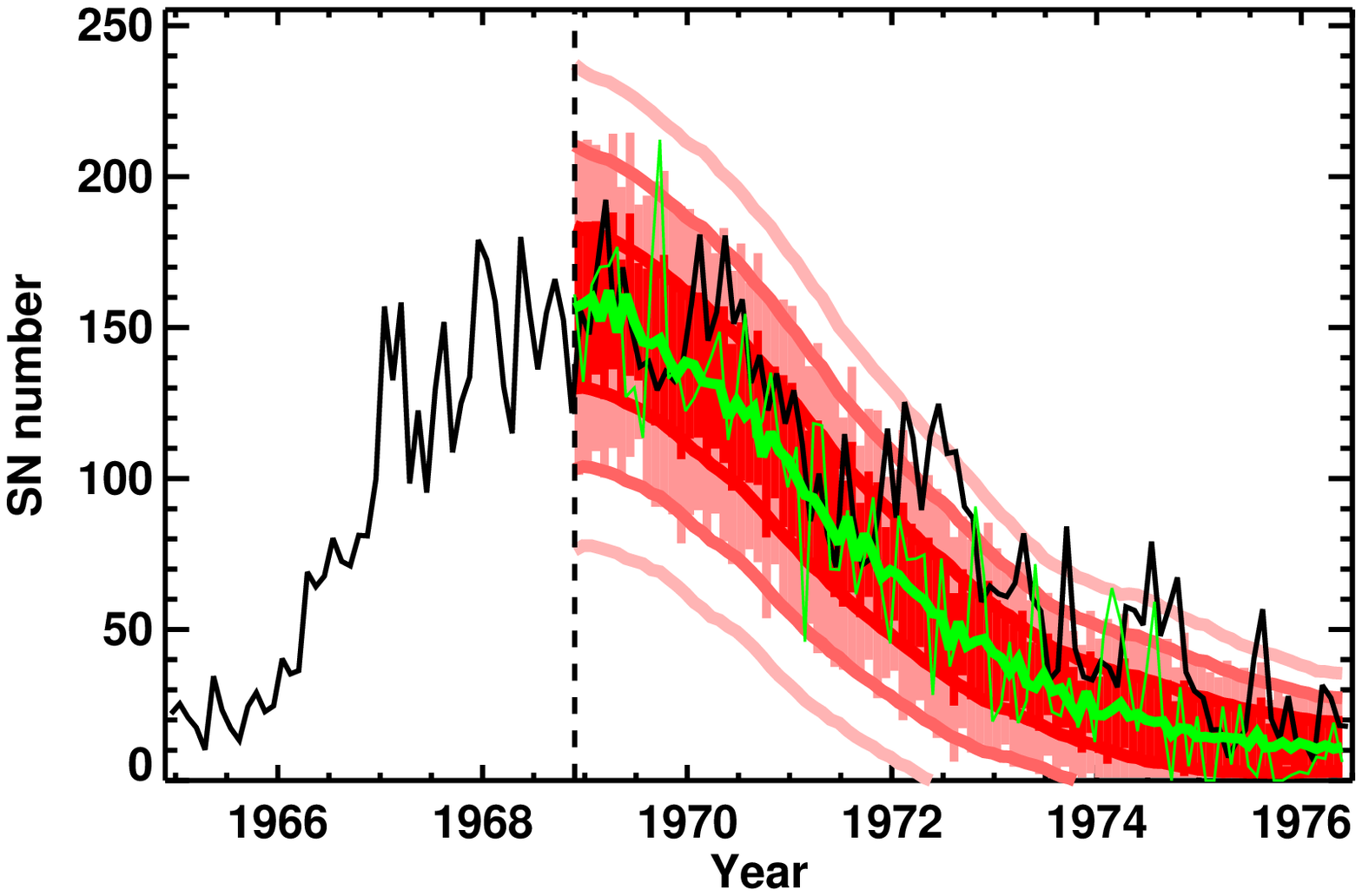}
  \includegraphics[width=0.32\textwidth]{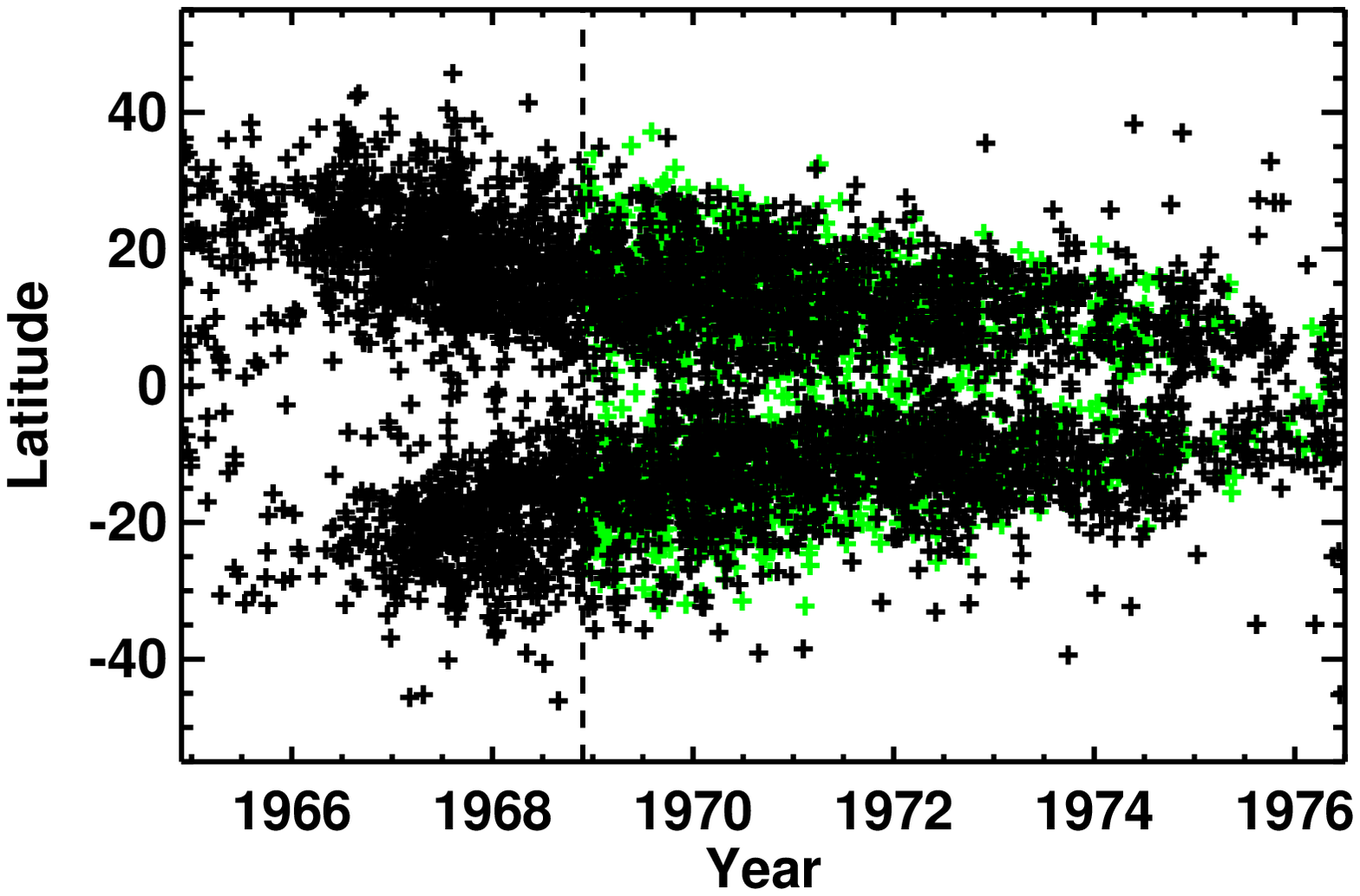}
  \includegraphics[width=0.32\textwidth]{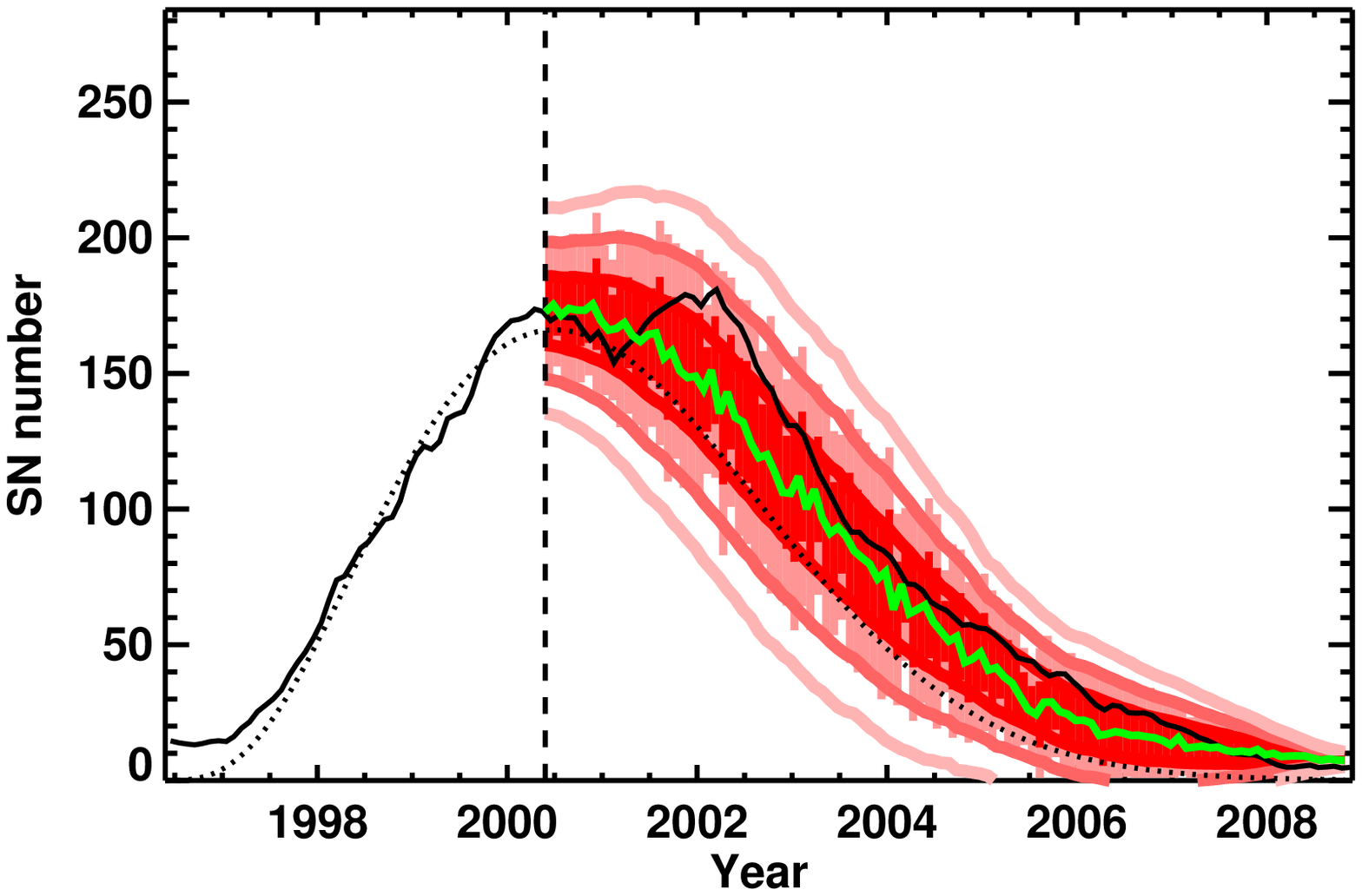}
  \includegraphics[width=0.32\textwidth]{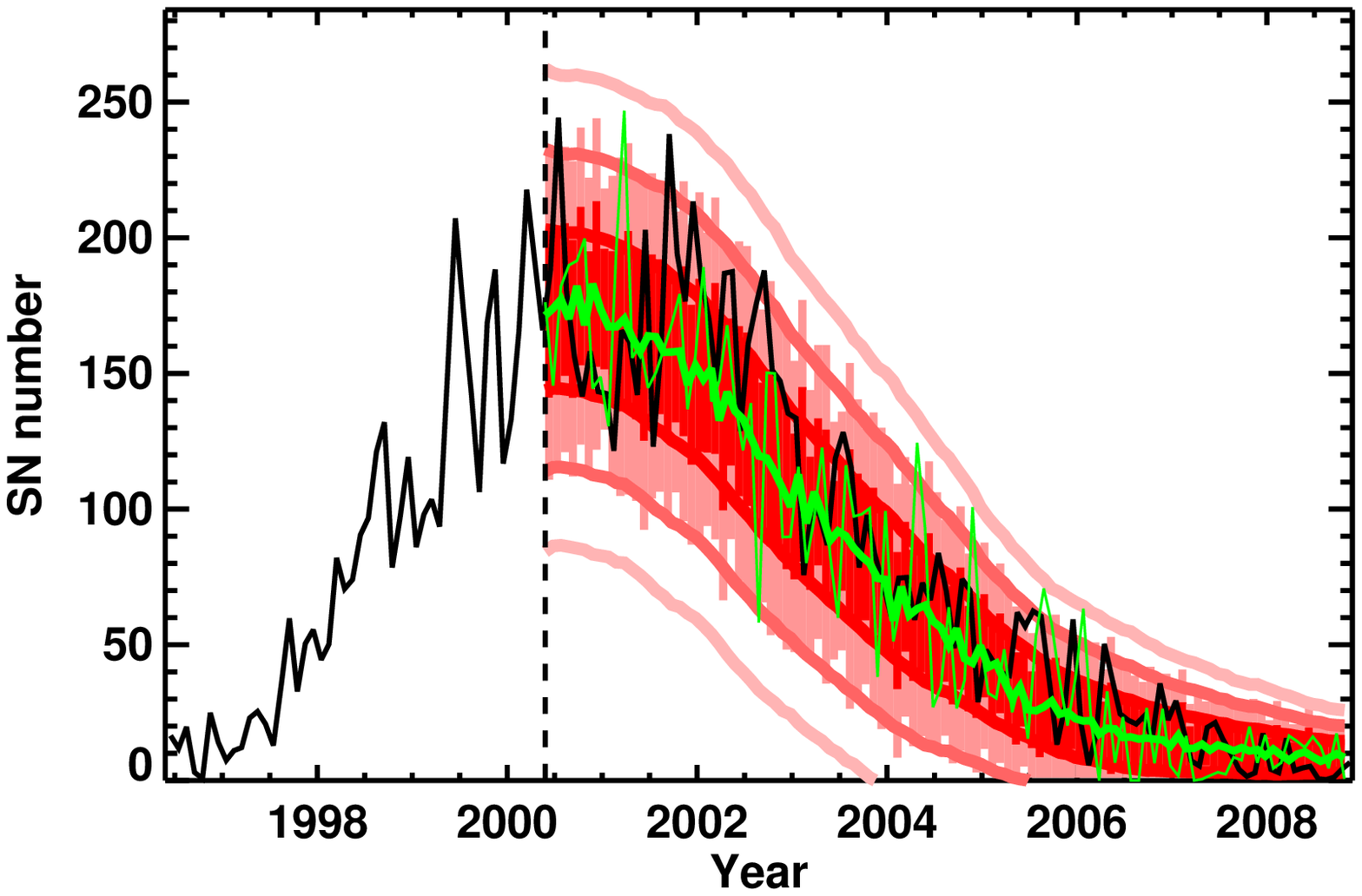}
  \includegraphics[width=0.32\textwidth]{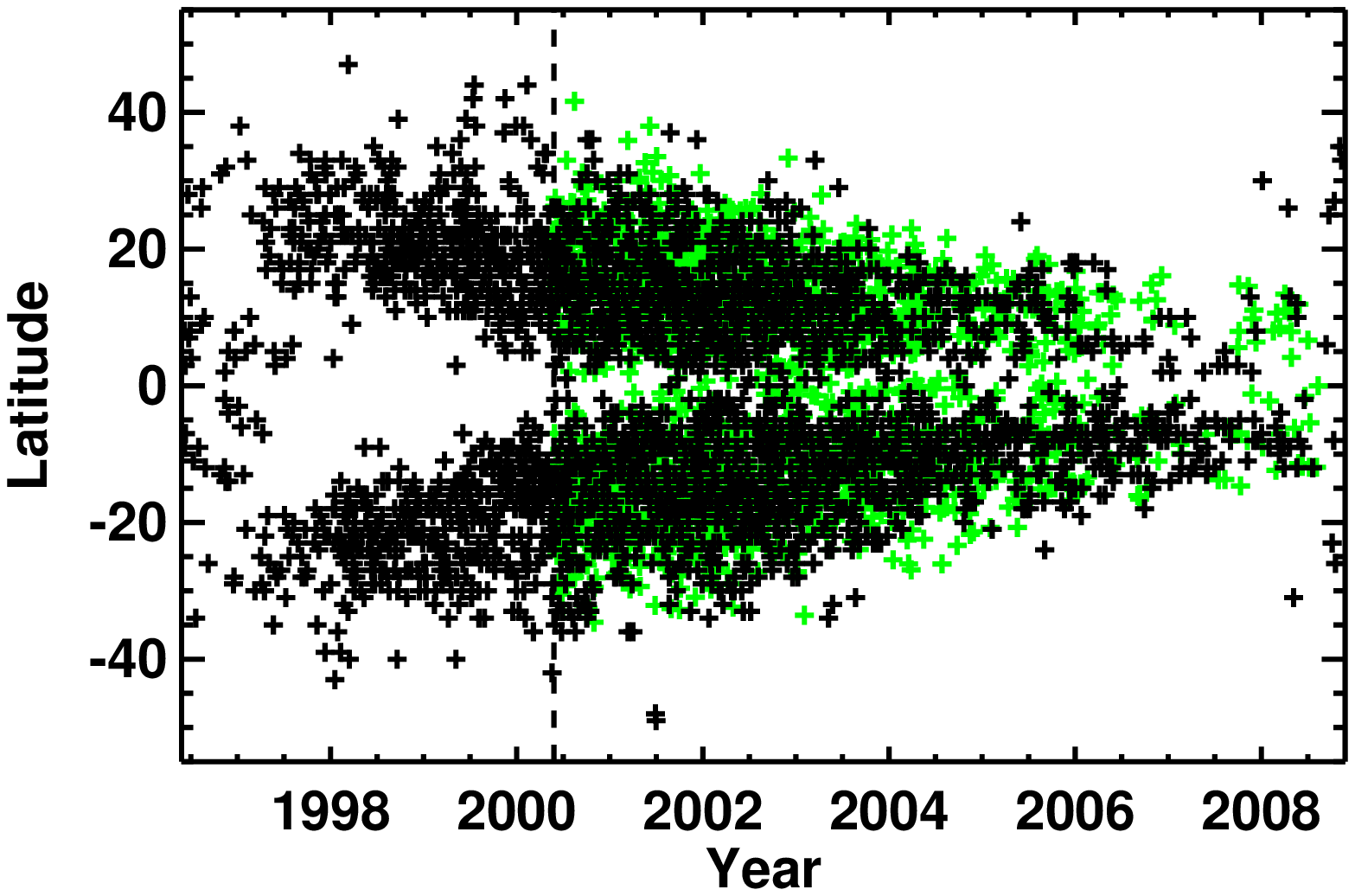}
  \includegraphics[width=0.32\textwidth]{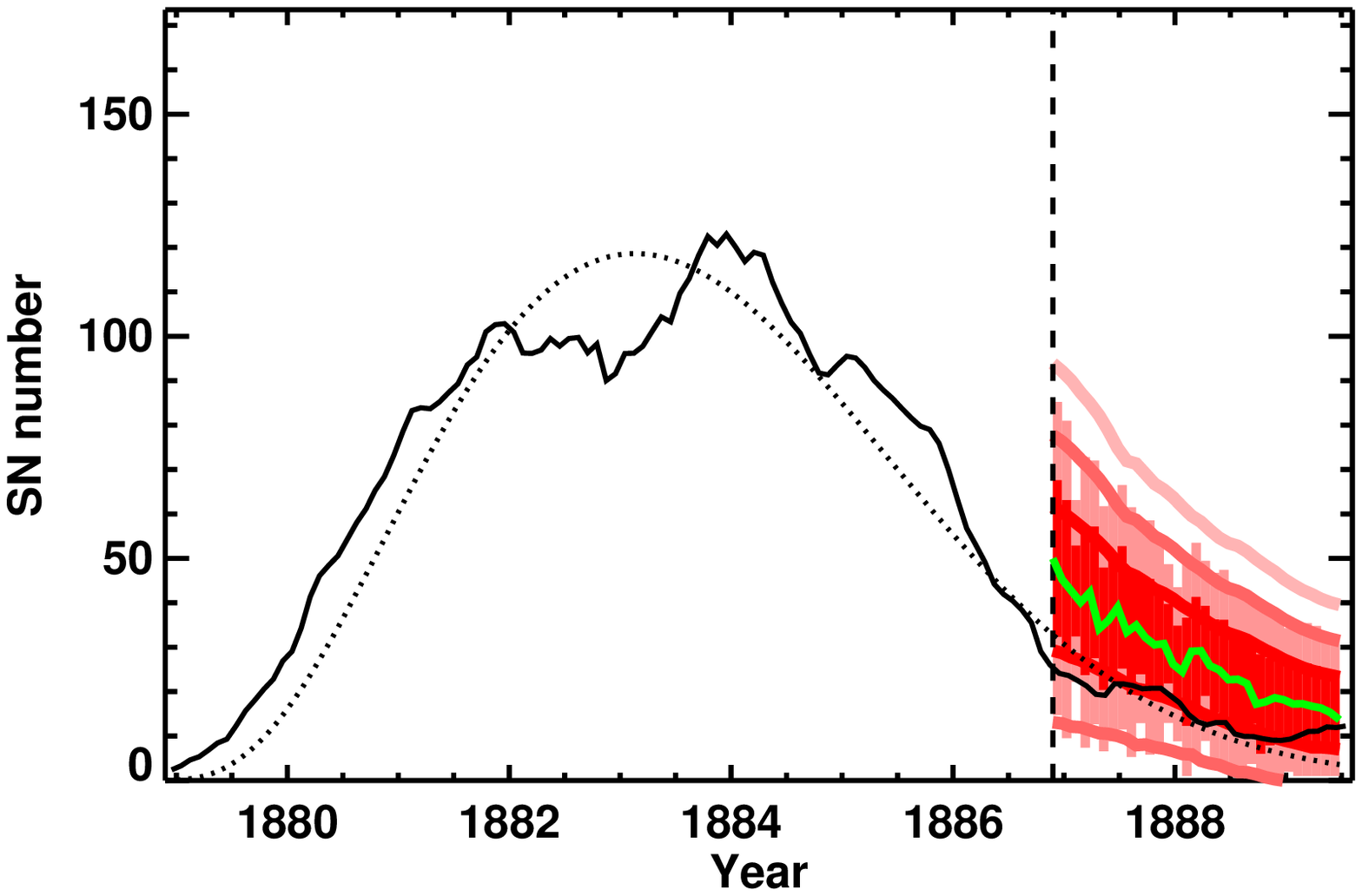}
  \includegraphics[width=0.32\textwidth]{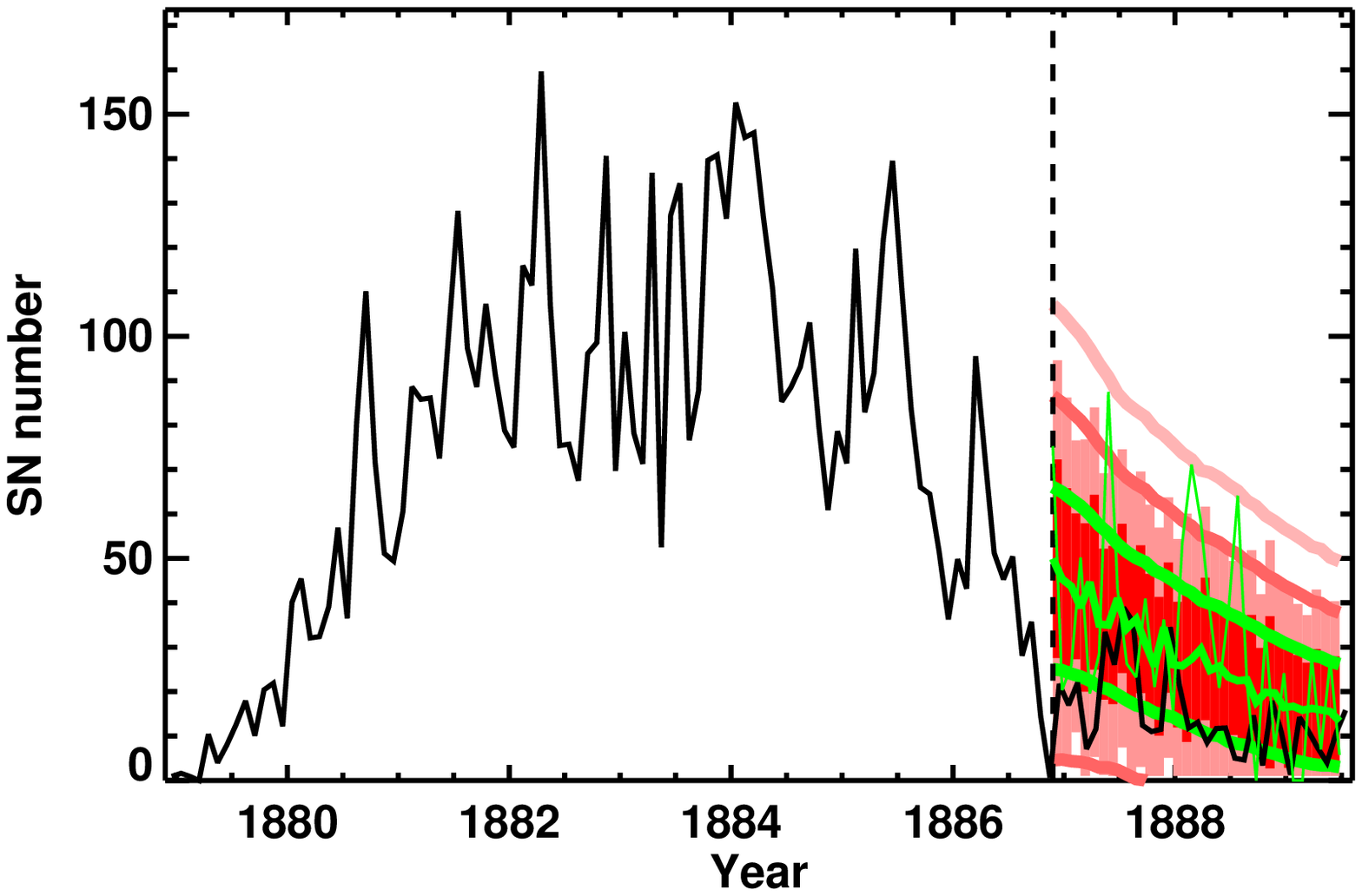}
  \includegraphics[width=0.32\textwidth]{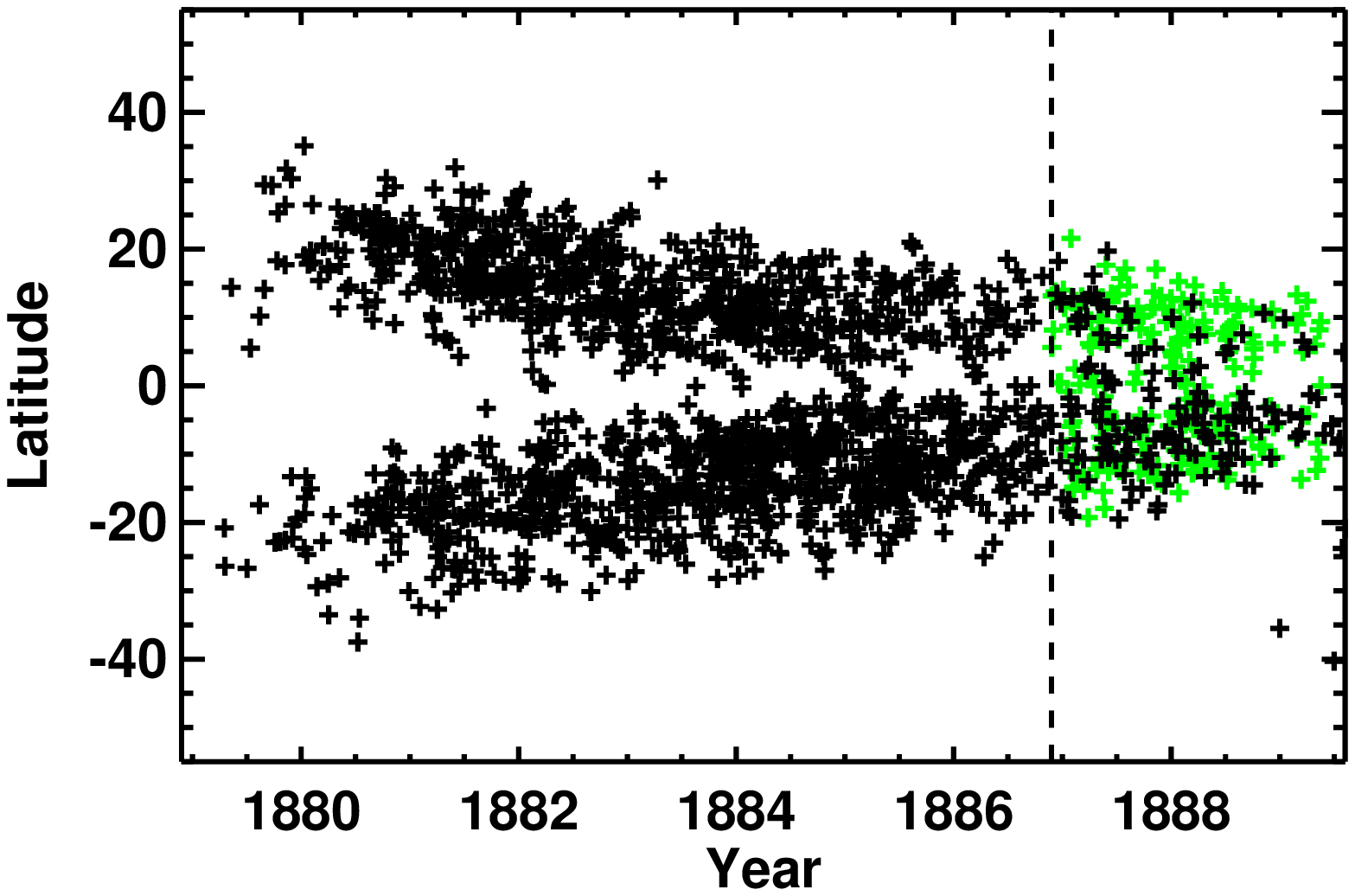}
  \includegraphics[width=0.32\textwidth]{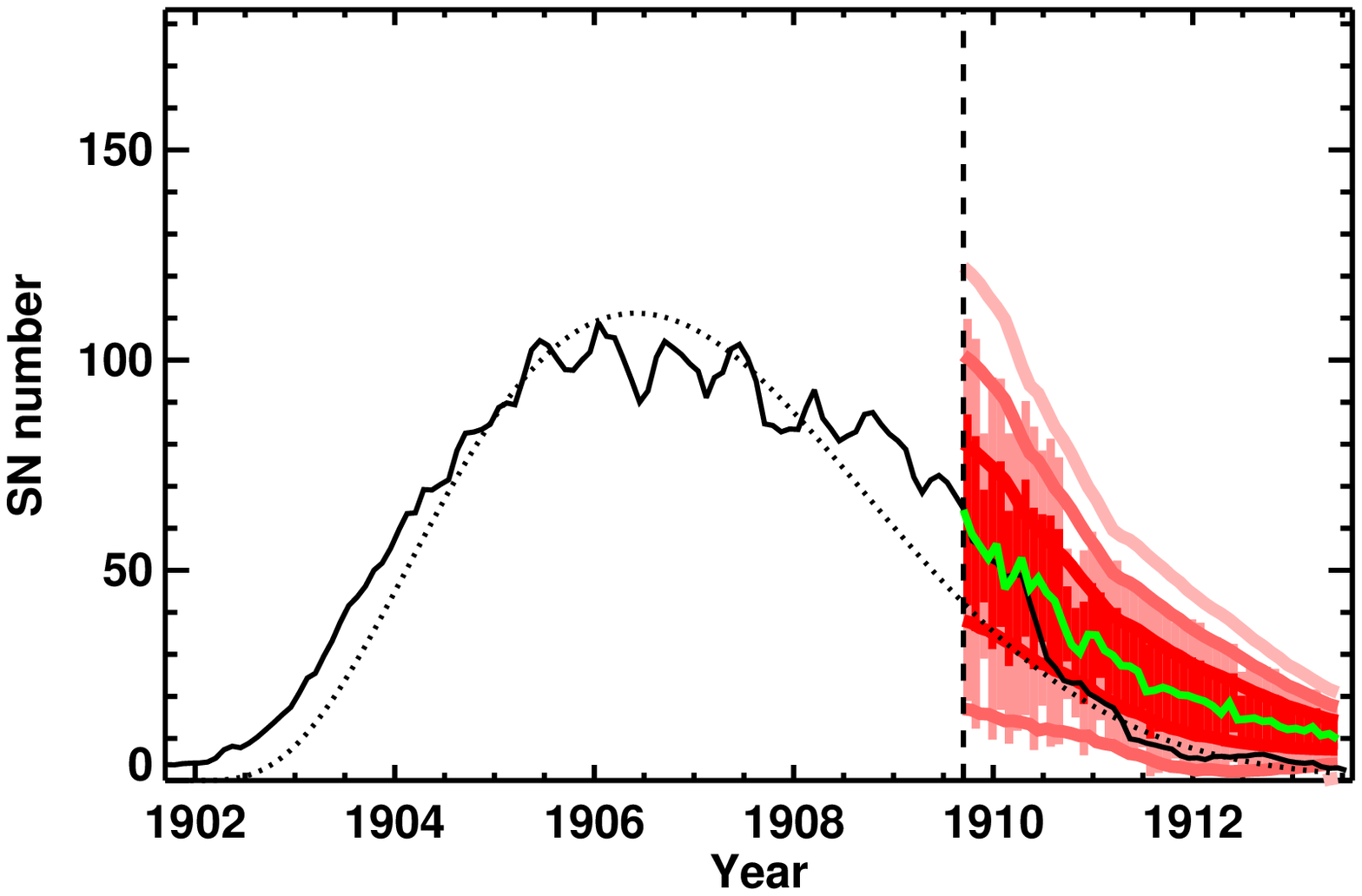}
  \includegraphics[width=0.32\textwidth]{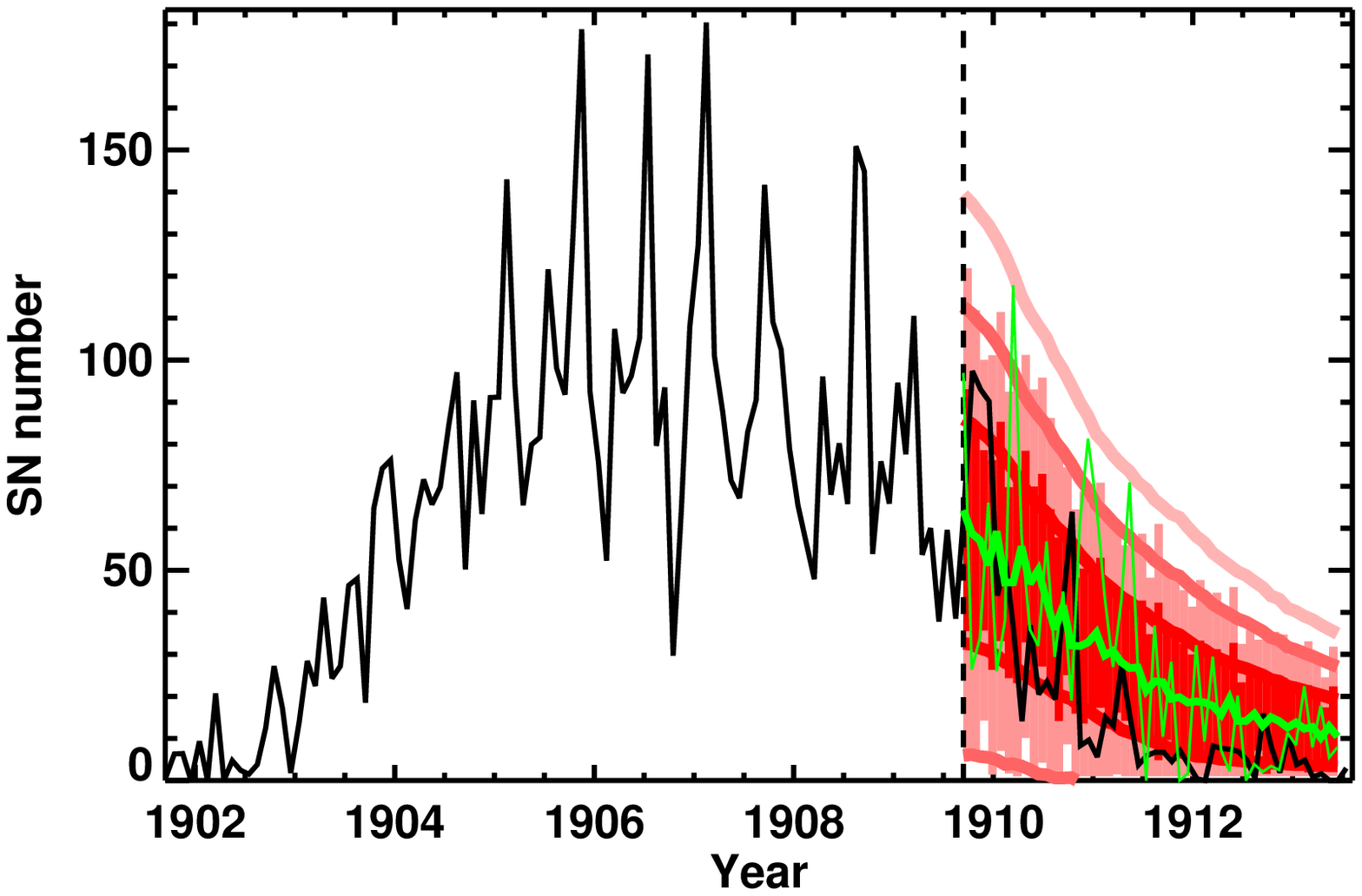}
  \includegraphics[width=0.32\textwidth]{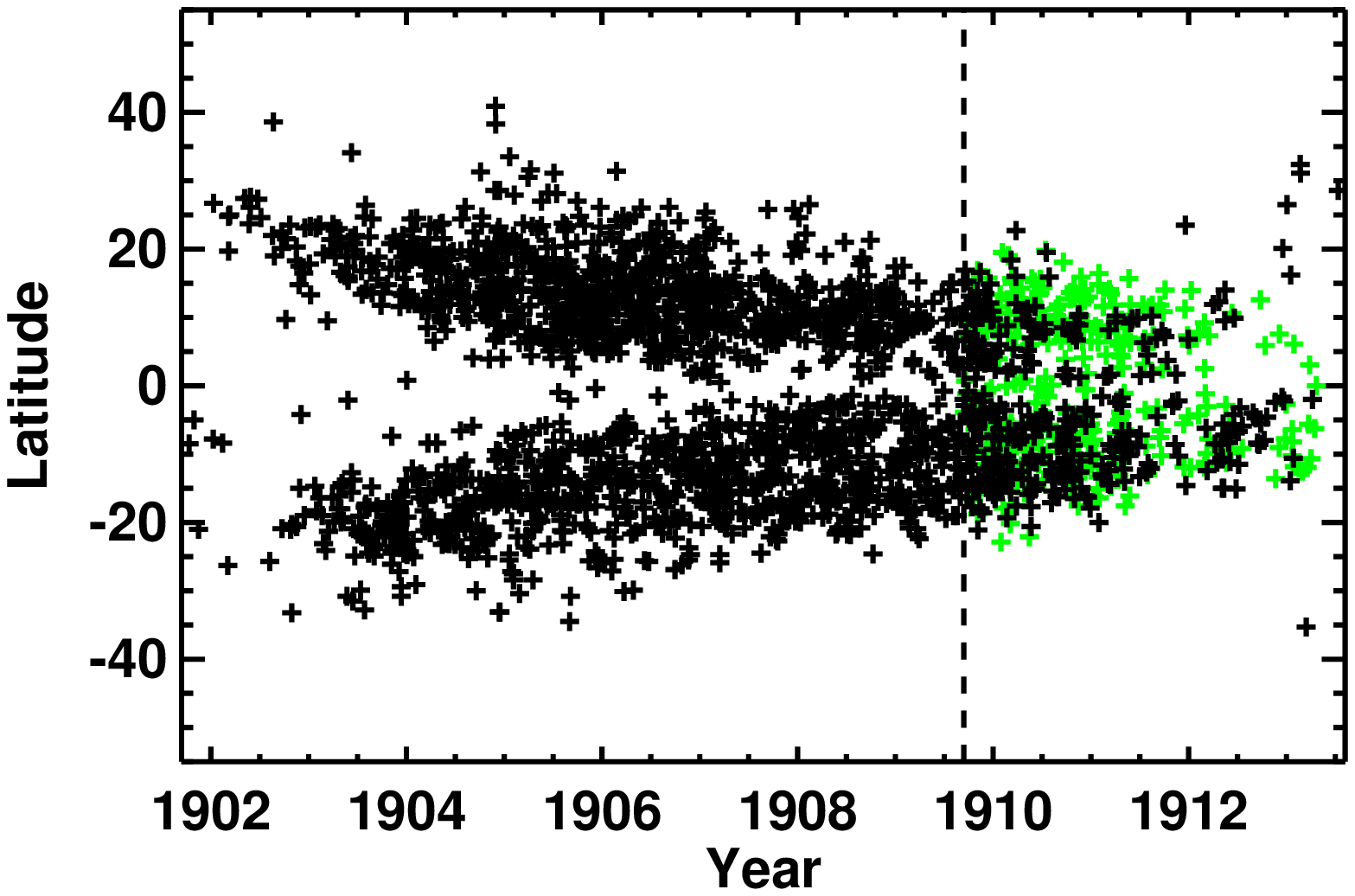}
  \caption{Examples for predicting the sunspot emergence of an ongoing solar cycle at different times after the start of the cycle. In the first column of all rows, the solid black curve shows the observed smoothed sunspot number. The other curves show fits and error bars from 4 years into cycle 20 (top row) and cycle 23 (second row to top), or 8 years into cycle 12 (third row) and cycle 14 (bottom row). The dotted black line shows the fit from Eq.(\ref{eq:cyFit}). The green solid curve is the predicted value using the scheme set out in this paper. The dark/light red shading gives the $\pm\sigma$/$\pm2\sigma$ variations of the predicted smoothed sunspot number. The three red curves below and above the mean values, respectively, show the boundaries of the $\pm\sigma$, $\pm2\sigma$ and $\pm3\sigma$ variations of the prediction. The second column is similar except it shows unsmoothed data and shows both the expectation value (thick green) as well as an example of one realization (thin green curve). The third column shows the observed the sunspot time-latitudinal distribution, i.e., butterfly diagrams in black, and one realization from the Monte-Carlo ensemble in green.}
\label{fig:PredictionExamples}
\end{figure}

Figure \ref{fig:PredictionExamples} shows the comparisons between the predicted and the observed sunspot number and the latitudinal location of sunspot groups (discussed in the next section). We take the following cases as examples: (1) cycle 20 (the first row) and cycle 23 (the second row) predicted at 4 years into cycles and showing smaller goodness-of-fits than HRW94, and (2) cycle 12 (the third row) and cycle 14 (the forth row) predicted at 8 years into the cycles and showing larger goodness-of-fits than HRW94. The first column is the smoothed monthly total sunspot number. Although the mean values of our predictions still always show the deviations from the observations, the observations are usually within our predicted 2$\sigma$ uncertainty ranges denoted by the shaded regions. This also demonstrates the importance of giving the uncertainty range. The second column is the time evolution of the monthly sunspot number. The random realizations look very similar, given the uncertainty limits, to the observed one.

\begin{table}[t]
\begin{center}
\caption{Measurements of the accuracy of the predicted mean values of the smoothed sunspot number evolution during cycles 12-24 at 4 years and 8 years into cycles by the goodness-of-fit $\chi$. The current method given by the paper and the profile given by HWR94 are both listed for comparisons.}
\begin{tabular}{c|cc|cc}
\tableline\tableline
 & \multicolumn{2}{c|}{Prediction at 4 yrs into a cycle}   &  \multicolumn{2}{c}{Prediction at 8 yrs into a cycle}  \\
 & current method & HWR94 & current method & HWR94 \\
 \tableline
cy.12 & 1.43& 1.94 & 1.10& 0.40\\
cy.13 & 0.73& 1.20 & 0.66& 1.71\\
cy.14 & 0.80& 1.20 & 1.32& 0.48\\
cy.15 & 1.02& 1.14 & 0.91& 0.44\\
cy.16 & 0.76& 1.16 & 0.88& 0.37\\
cy.17 & 0.79& 1.04 & 0.37& 1.39\\
cy.18 & 0.61& 1.27 & 0.13& 1.07\\
cy.19 & 0.83& 1.21 & 0.78& 1.87\\
cy.20 & 1.83& 2.91 & 1.53& 2.85\\
cy.21 & 0.65& 1.01 & 0.70& 0.64\\
cy.22 & 0.61& 0.99 & 0.51& 0.84\\
cy.23 & 0.84& 1.90 & 0.46& 1.59\\
cy.24 & 1.18& 1.01 & -- & -- \\
\tableline\tableline
\label{tab:GoodnessOfFit}
\end{tabular}
\end{center}
\end{table}

\begin{table}[t]
\begin{center}
\caption{Quantitative measurements of the prediction of the 13-month smoothed monthly sunspot number $F_{\textrm{sm}}(t)$. The values show the percentages of the observed sunspot number, which are out of the 1$\sigma$, 2$\sigma$ and 3$\sigma$ of the predicted variations.}
\begin{tabular}{c|ccc|ccc}
\tableline\tableline
 & & Prediction at 4 yrs into a cycle  &   &  & Prediction at 8 yrs into a cycle &  \\
 & $1\sigma$ & $2\sigma$ & $3\sigma$ & $1\sigma$ & $2\sigma$ & $3\sigma$\\
 \tableline
cy.12 & 47.5& 11.2 & 0.0 &43.7 &0.0 &0.0 \\
cy.13 & 0.0 & 0.0 & 0.0& 12.2& 0.0& 0.0\\
cy.14 & 6.4 & 0.0 & 0.0& 54.3& 0.0 & 0.0\\
cy.15 & 20.8& 5.5 & 0.0& 29.1& 0.0& 0.0\\
cy.16 & 14.9& 0.0& 0.0& 15.4& 0.0& 0.0\\
cy.17 & 18.4& 0.0& 0.0& 0.0& 0.0& 0.0\\
cy.18 & 8.2 & 0.0& 0.0& 12.0& 0.0& 0.0 \\
cy.19 & 17.7& 0.0& 0.0& 4.0& 0.0& 0.0\\
cy.20 & 70.3& 37.3& 0.0& 67.4& 13.9& 0.0\\
cy.21 & 13.3& 0.0&0.0 & 14.8&0.0 &0.0 \\
cy.22 & 5.9& 0.0& 0.0& 0.0&0.0 &0.0 \\
cy.23 & 20.6& 0.0 & 0.0 & 0.0 &0.0 & 0.0\\
cy.24 & 39.5& 10.4 &0.0 &  & & \\
\tableline\tableline
\label{tab:phases_rstd_sm}
\end{tabular}
\end{center}
\end{table}

\begin{table}[t]
\begin{center}
\caption{Quantitative measurements of the prediction of the monthly total sunspot number $F_{\textrm{mn}}(t)$. The values show the percentages of the observed sunspot number, which are out of the 1$\sigma$, 2$\sigma$ and 3$\sigma$ of the predicted variations.}
\begin{tabular}{c|ccc|ccc}
\tableline\tableline
 & & Prediction at 4 yrs into a cycle  &   &  & Prediction at 8 yrs into a cycle &  \\
 & $1\sigma$ & $2\sigma$ & $3\sigma$ & $1\sigma$ & $2\sigma$ & $3\sigma$\\
 \tableline
cy.12 & 47.5& 17.5 & 8.7 &43.7 &0.0 &0.0 \\
cy.13 & 29.8& 3.1 & 0.0& 40.8& 0.0& 0.0\\
cy.14 & 32.0& 16.0& 9.3& 17.4& 0.0 & 0.0\\
cy.15 & 40.3& 9.7& 5.5& 50.0& 4.2& 4.2\\
cy.16 & 39.2& 5.4& 1.4& 34.6& 0.0& 0.0\\
cy.17 & 39.4& 5.3& 0.0& 28.6& 3.6& 0.0\\
cy.18 & 28.8& 6.8& 0.0& 28.0& 0.0& 0.0 \\
cy.19 & 25.3& 3.8& 0.0& 24.0& 0.0& 0.0\\
cy.20 & 46.1& 24.2& 7.7& 34.9& 16.3& 7.0\\
cy.21 & 38.7& 2.7&0.0 & 37.0&0.0 &0.0 \\
cy.22 & 31.3& 3.0& 0.0& 21.0&0.0 &0.0 \\
cy.23 & 31.4& 5.9& 0.0& 16.7 &0.0 &0.0\\
cy.24 & 35.4& 8.3& 0.0 & & & \\
\tableline\tableline
\label{tab:phases_rstd}
\end{tabular}
\end{center}
\end{table}

\subsection{Prediction of the BMR emergence}
Once we have the predicted sunspot number, the next step is to predict the emergence properties of the sunspot groups. \cite{Jiang2011} give the dependence of the statistical properties of sunspot group emergence, including the latitude, longitude, area and tilt angle of sunspot groups in the form of the BMRs on the cycle phase and strength. Those properties are used here. The following list summarizes the statistical properties aiming to convert the sunspot number time series into the time series of sunspot group emergence.

\begin{itemize}
\item The number of BMRs emerging per month was taken to be equal to $F_{BMR}(t)=0.24F_{\textrm{mn}}(t)$. The number of the daily emergence of the BMR is randomly realized satisfying the total number $F_{BMR}(t)$.
\item The mean latitudinal distribution of BMR emergence $\lambda_n$ depends on the cycle strength $S_n$. It obeys the form of $\lambda_n(x)=(26.4-34.2x+16.1x^2)(\overline{\lambda}_n/\langle\lambda_n\rangle_{12-20})$, where $\overline{\lambda}_n=(12.2+0.015S_n)/14.6$ and $x$ is the fraction of a solar cycle period.
\item The width of the latitude distribution obeys a Gaussian profile with a half width of $\sigma$, which is equal to $(0.14+1.05x-0.78x^{2})\lambda_n(x)$. We exclude points deviating from the mean by more than 2.2 $\sigma$ for the equatorward side.
\item The BMR emergence has the symmetric distribution in north and south hemispheres.
\item The BMR emergence has the random longitudinal distribution.
\item The number density function of sunspot group areas obey the following distributions,
\begin{equation}
n_{BMR}(A_S)=\begin{cases}
0.3A_S^{-1.1} & \mathrm{for~} A_S \leq 60 \mu \mathrm{Hem} \\
0.003\exp[-\frac{1}{2\ln 3}(\ln A_S-\ln 45)^2] & \mathrm{for~} A_S > 60\mu \mathrm{Hem}.
\end{cases}
\end{equation}
\item The average areas of sunspot groups are cycle phase ($x$) dependent, which obeys $\overline{A_S}(x)=115+396x-426x^2$.
\item The umbra area of each BMR is $A_U=A_S/5$. The total area of BMR $A=A_S+A_P$, where $A_P$ is the plage area and is calculated by $A_P=414+21A_S-0.0036A_S$ \citep{Chapman1997}.
\item The mean tilt angle, $\overline{\alpha_n}(\lambda)$, of the emerging BMRs for the cycle $n$ is assumed to follow Joy's law in the form
$\overline{\alpha_n}=T_n\sqrt{|\lambda|}$, where $T_n=1.72-0.0022S_n$.
\item For the scatter of the tilt angle, $\Delta\alpha$, its averaged value is zero and the standard deviation is $\sigma_\alpha$. We use the empirical relation between $\sigma_\alpha$ and $A_U$, which is $\sigma_\alpha=-11\log A_U+35$  to derive $\sigma_\alpha$ \citep{Jiang2014b}.
\item A factor 0.7 is multiplied to the resulting tilt angle $\alpha_n=\overline{\alpha_n}+\sigma_\alpha$ to account for the effect of latitudinal inflow towards BMRs \citep{Jiang2010, Martin-Belda2017}.
\end{itemize}

The third column of Figure \ref{fig:PredictionExamples} shows examples of the time evolution of the sunspot latitudinal distribution, i.e., butterfly diagrams. The predicted one is consistent with the observed one.

The BMR emergence provides the source of the large-scale magnetic field over the solar surface. The evolution of the large-scale magnetic field can be described by the SFT model, which will be presented in the following section.

\section{Prediction of the large-scale magnetic field evolution}
\subsection{Surface flux transport model}
The relevant equation to describe the evolution of the large-scale magnetic flux distribution at the solar surface $B(\lambda,\phi,t)$ as a combined result of the emergence of BMRs, a random walk due to supergranular flows, and the transport by large-scale surface flows is as follows.
\begin{eqnarray}
\label{eqn:SFT} \nonumber\frac{\partial B}{\partial t}=& &
-\Omega(\lambda)\frac{\partial B}{\partial \phi}
-\frac{1}{R_\odot\cos\lambda}\frac{\partial}{\partial \lambda}[\upsilon(\lambda)B\cos\lambda] \\
\noalign{\vskip 2mm}
& & +\eta \left[\frac{1}{R_\odot^2\cos{\lambda}}
\frac{\partial}{\partial \lambda}\left(\cos\lambda\frac{\partial B}{\partial \lambda}\right) +
\frac{1}{R_\odot^2 \cos^2{\lambda}}\frac{\partial^2 B}{\partial\phi^2}\right] + S(\lambda,\phi,t),
\end{eqnarray}
where $\lambda$ and $\phi$ are heliographic latitude and longitude, respectively. The latitudinal differential rotation of Sun, $\Omega(\lambda)$, is taken from \citet{Snodgrass1983} and the poleward meridional flow, $\upsilon(\lambda)$, is given by \citet{Ballegooijen1998}. It is consistent with the measurement by \cite{Hathaway2011}. The magnetic diffusivity $\eta$ is 250 km$^2$s$^{-1}$, which is within the range from the observational studies summarized by \cite{Schrijver1996}. All the three transport parameters are time independent. $S(\lambda,\phi,t)$ is the time dependent flux source term, which will be obtained based on the predicted sunspot emergence using the empirical method in Section \ref{sec:PrdcEmergence}. The initial magnetic field configuration due to the BMR emergence based on its area, location and tilt is given by \cite{Baumann2004}. The corresponding magnetic flux is determined by a single parameter, Bmax (=592G), which is calibrated by the total unsigned surface flux obtained from SOHO/MDI polar field corrected synoptic maps \citep{Sun2011} after rebinning to the spatial resolution of the simulation ($1^\circ$ in both latitude and longitude).

\begin{figure}[!htp]
 \centering
\includegraphics[scale=0.3]{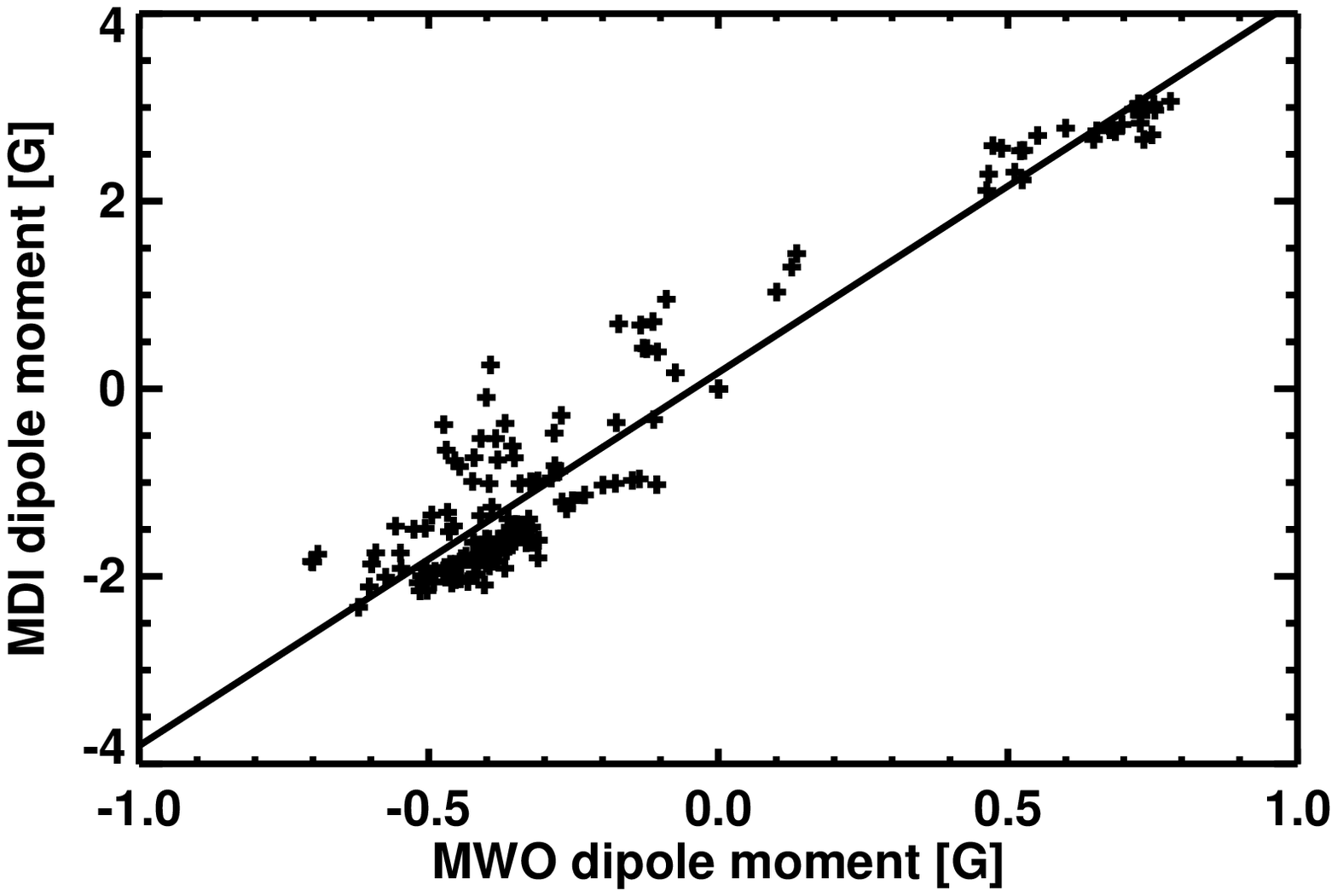}
\includegraphics[scale=0.3]{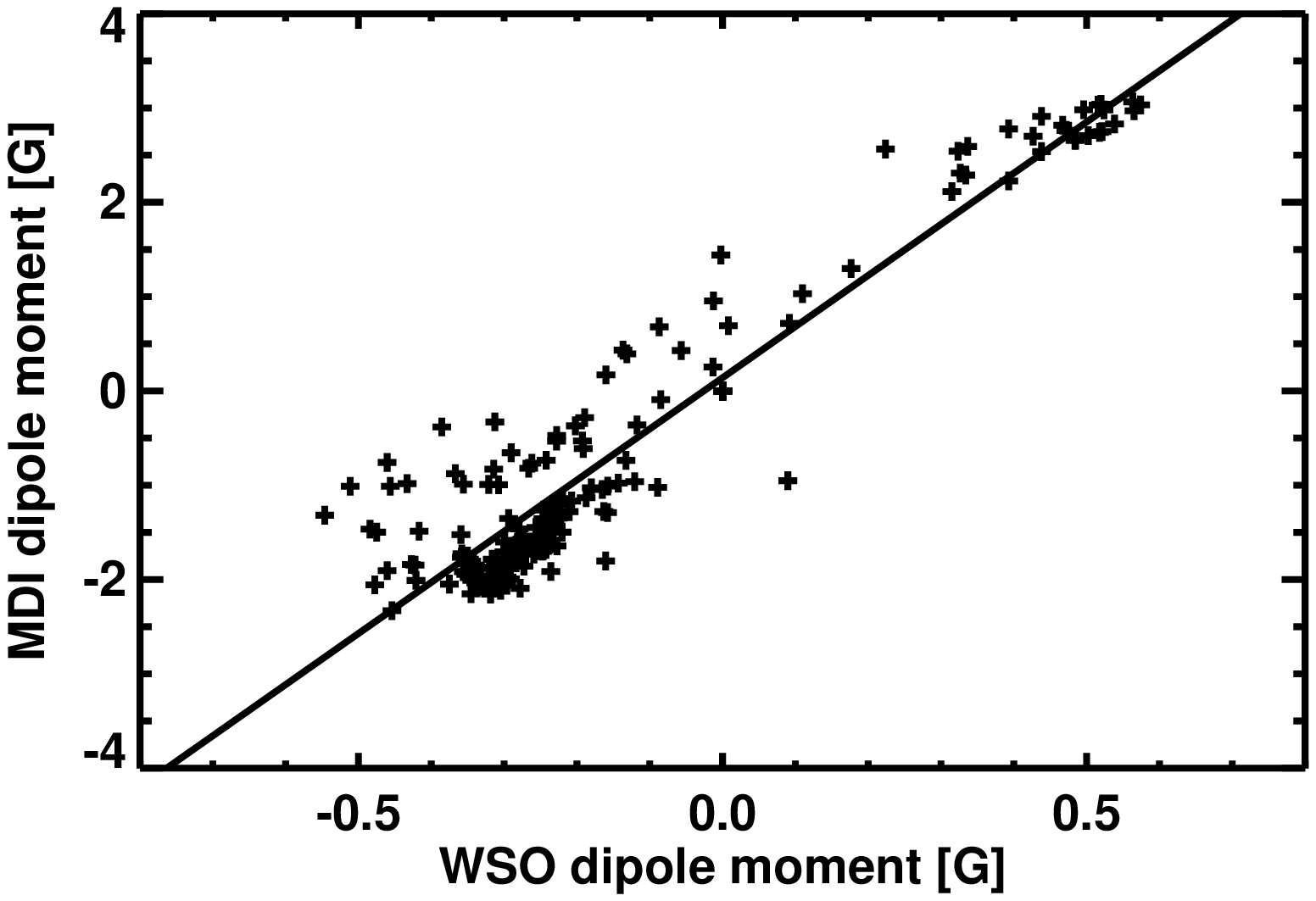}
\includegraphics[scale=0.3]{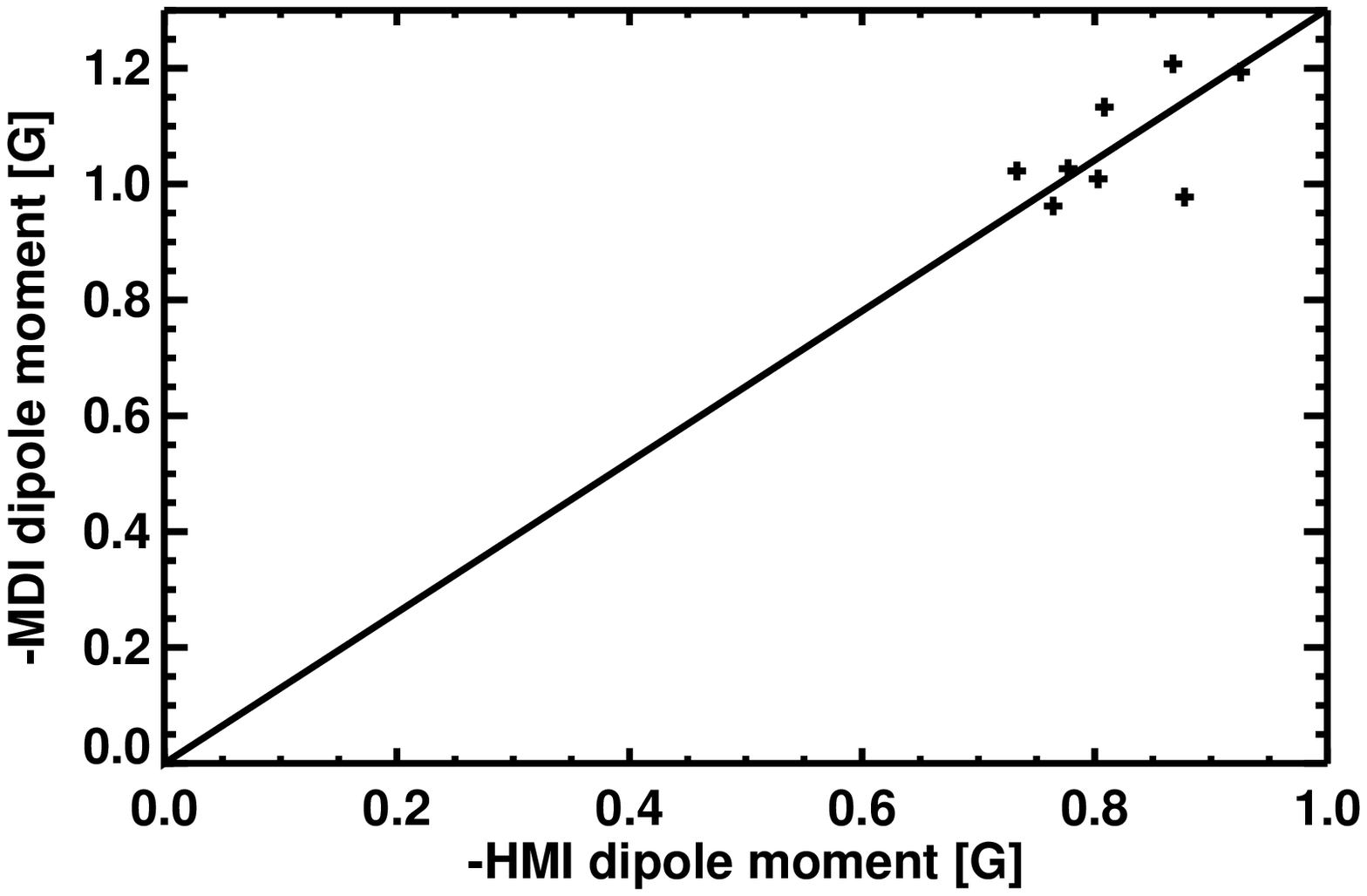}
\caption{Cross-calibrations of the axial dipole moments from different observed synoptic magnetograms. Left panel: MWO verse MDI ($D_{\rm{MDI}}=0.17+3.98D_{\rm{MWO}}$, correlation coefficient $r=0.94$); Middle panel: WSO verse MDI ($D_{\rm{MDI}}=0.14+5.42D_{\rm{WSO}}$, $r=0.96$); Left panel: HMI verse MDI ($D_{\rm{MDI}}=1.30D_{\rm{HMI}}$, $r=0.89$).}
\label{fig:DMs_calibrations}
\end{figure}

For the initial condition of each SFT simulation, we use the SOHO/MDI polar field corrected synoptic maps (available from 1996 June until 2010 May) and the radial synoptic maps with 3600 points in Carrington longitude and 1440 points equally spaced in sine latitude from HMI (2010 May until the present).
The abnormal points which correspond to unobserved values or the absolute values over 50G above $\pm60^{\circ}$ latitudes are filled by the averaged value of neighbouring normal 3 points. Both the MDI and the HMI data are smoothed with the width of 7. Both data were reduced to a resolution of $1^\circ$ in latitude and longitude by the IDL CONGRID function and then were converted to the equal latitudes. Rightmost of Figure \ref{fig:DMs_calibrations} shows that the cross-calibration of HMI and MDI based on the axial dipole moment during the overlap time period (Carrington Rotations, CRs2097-2104) is 1.3. This is consistent with \cite{Liu2012} who compared Line-of-Sight magnetograms taken by MDI and HMI. We multiply a factor 1.3 when the HMI synoptic map is used as the initial condition. The low resolution data of synoptic maps were used by \citet{Cameron2016} and \citet{Jiang2017}.

We used the code originally developed by \cite{Baumann2004} to do the numerical calculations. The magnetic field is expressed in terms of spherical harmonics up to $l$ = 63. A fourth-order Runge-Kutta method is used for time stepping with the interval of one day.

\subsection{Prediction of the large-scale magnetic field evolution over surface}
With 50 sets of random realizations of the magnetic flux emergence and the synoptic magnetograms as the initial condition, we may derive the possible large-scale magnetic field evolution over the surface using the SFT model. Since here our interest mainly concentrates on long-term predictions, we mainly show the results about the time evolution of the axial dipole moment $D(t)$, which is defined as
\begin{equation}
\label{eq:dipole}
D(t)=\frac{3}{2} \int_0^{180} \left\langle B\right\rangle(\theta,t)
              \cos\theta\sin\theta d\theta.
\end{equation}
The uncertainties of the results originate from two ingredients. One is the uncertainty due to the scatter in the properties of the BMRs emergence, which are assumed to be reasonably represented in Sections \ref{sec:CycleShape} and \ref{sec:PrdcEmergence}. The other is the uncertainty due to measurement error in the magnetograms used as initial conditions. The imperfect measurements mainly result from the instrumental characteristics, e.g., spatial resolution, scattered light, and filter characteristics and from the inherent complexity of the solar magnetic field, e.g., the saturated factors for different spectral lines \citep{Wang1995}. The properties of the magnetic field at lower latitudes have larger effects on the results \citep{Jiang2014b}. Since some imperfections of the measurements are inevitable, they are potential uncertainties to the predictive skill of the model. Here we only consider the averaged radial surface field over the surface, which deviates from zero due to measurement error. The same method described in \cite{Cameron2016b} to estimate the error is used. The total error due to the magnetogram measurement and the random flux source emergence is determined by adding them quadratically.

\begin{figure}[!htp]
 \centering
\includegraphics[scale=0.4]{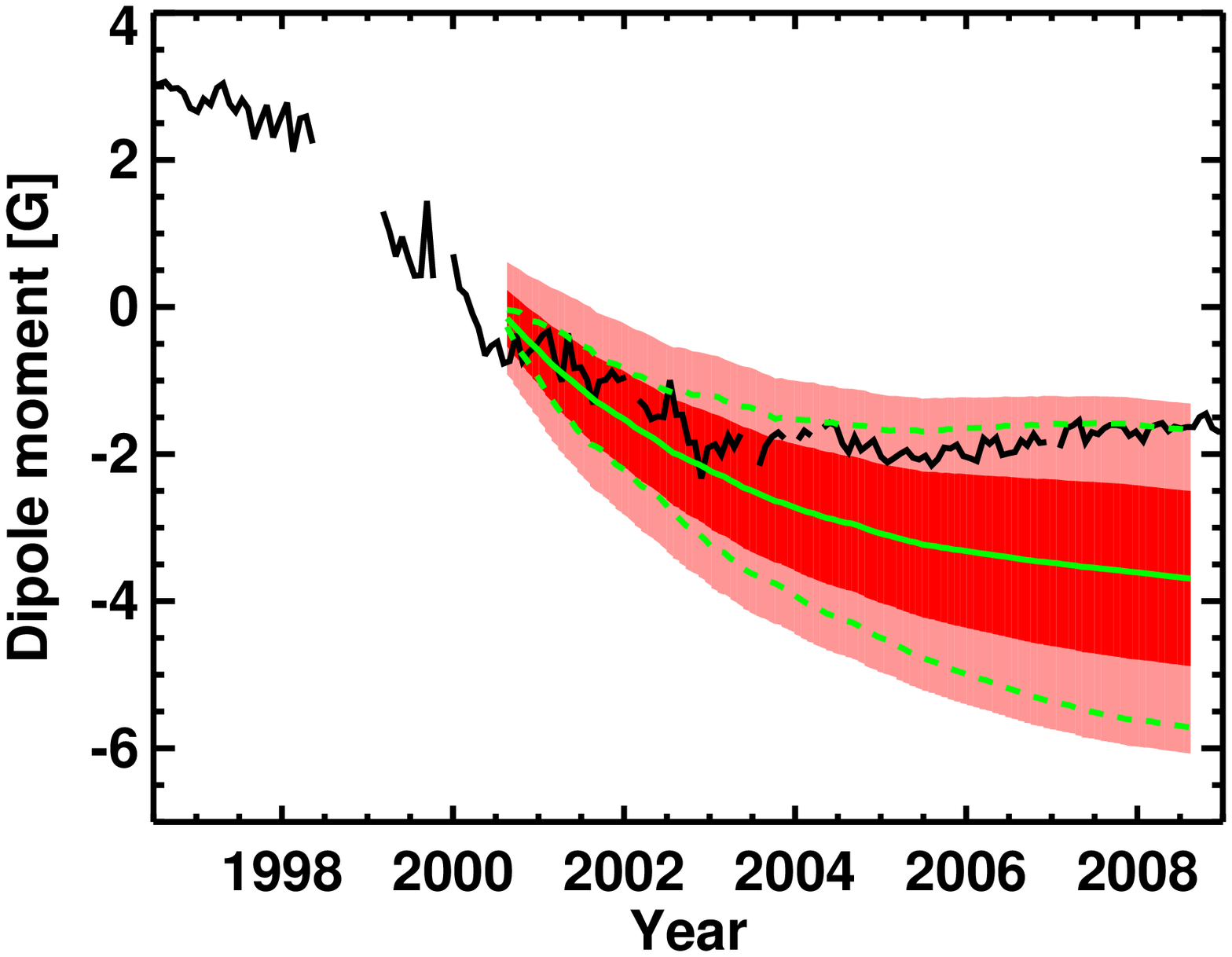}
\includegraphics[scale=0.4]{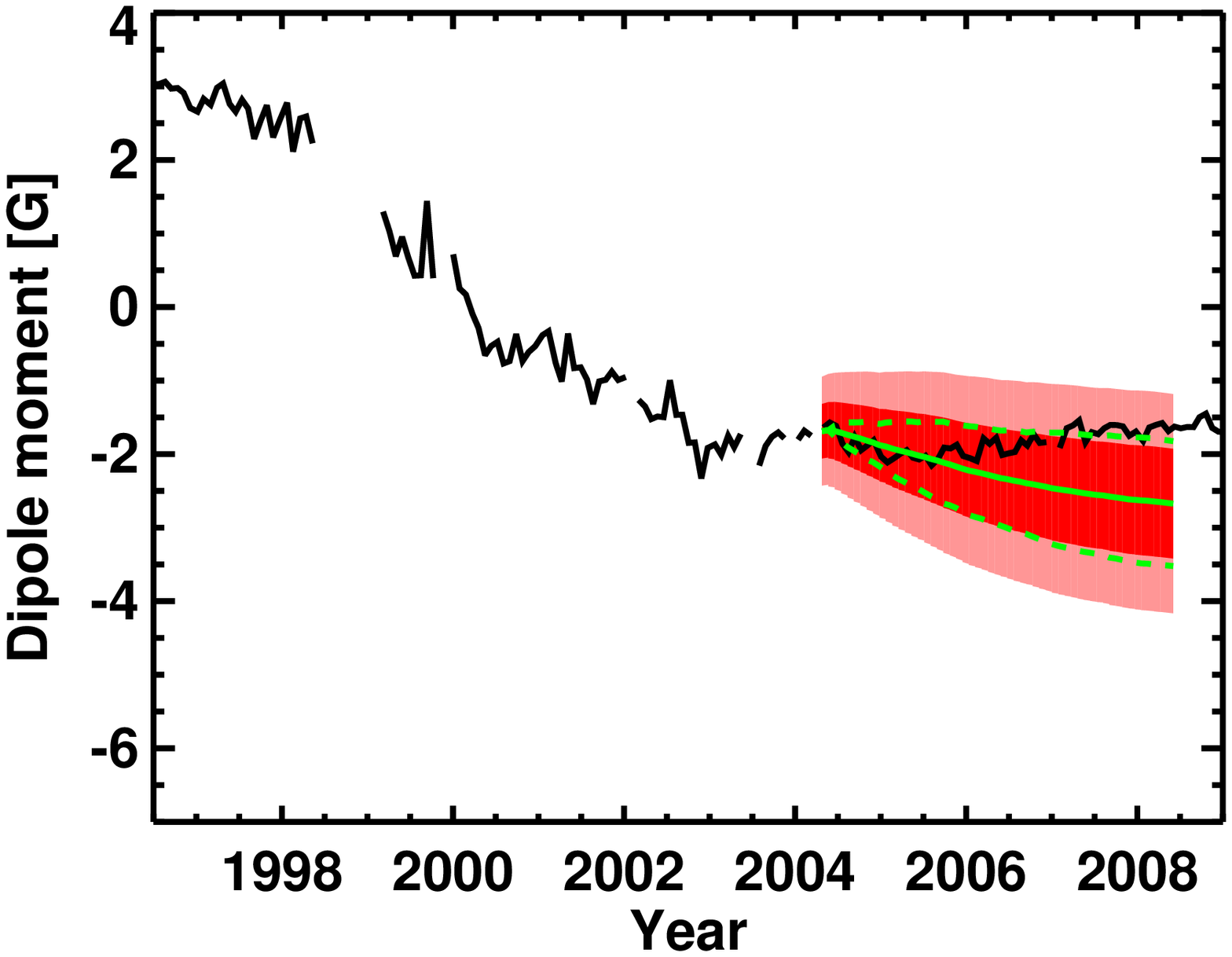}
\includegraphics[scale=0.4]{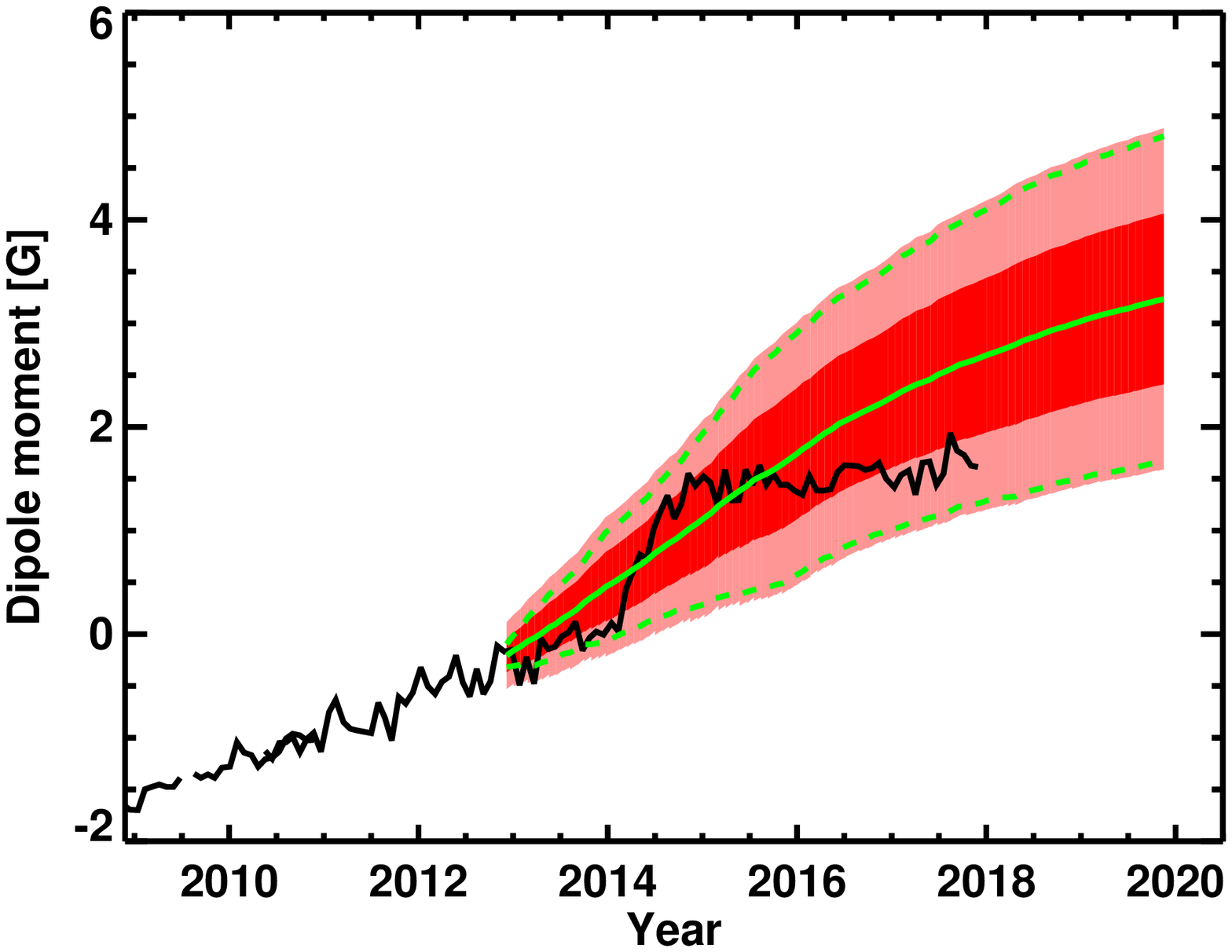}
\includegraphics[scale=0.4]{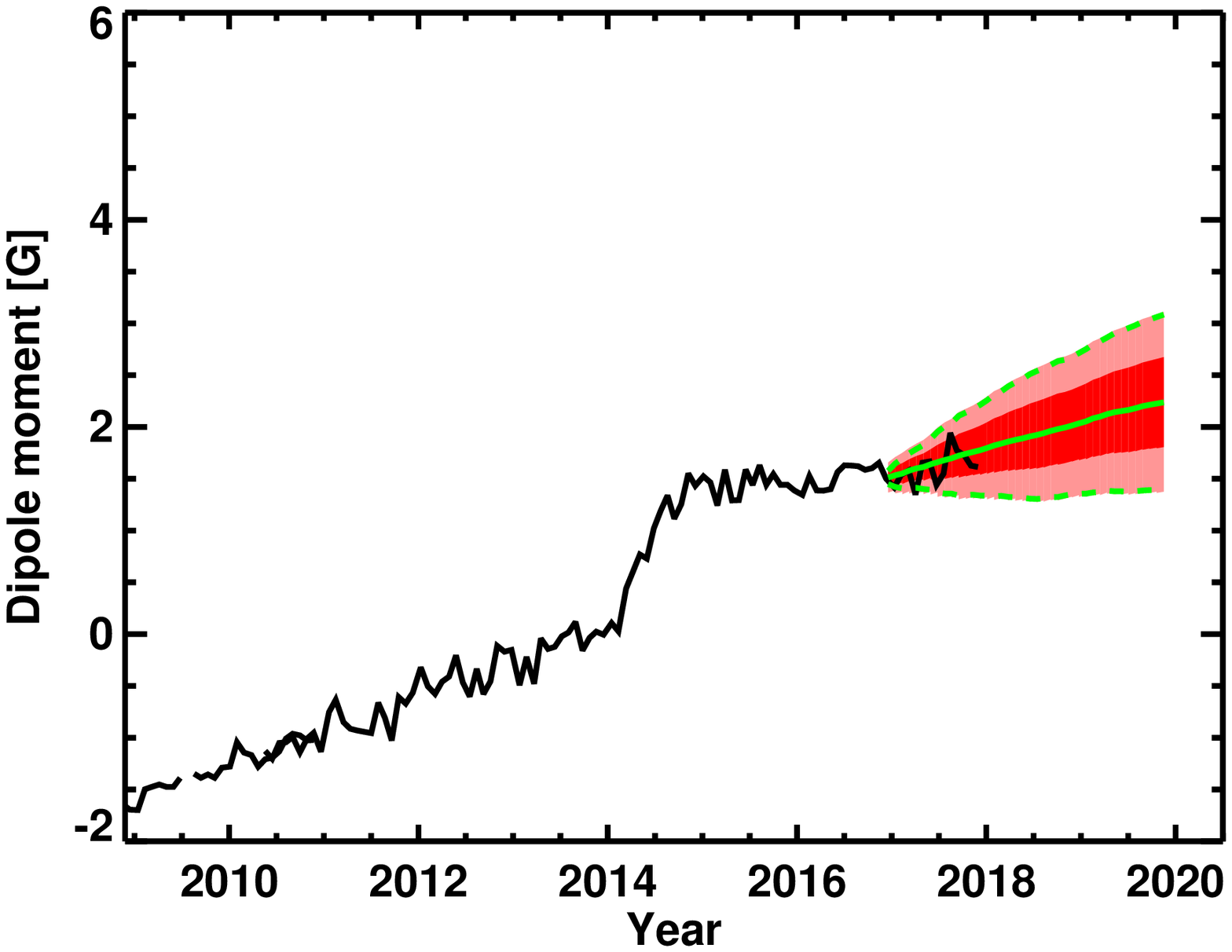}
\caption{Examples of the predicted time evolutions of the dipole moments for cycle 23 (upper panels) and cycle 24 (lower panels) when 4 years into each cycle (left panels) and 8 years into each cycle (right panels). Solid green lines show the averages of 50 SFT simulations with random sources starting from the prediction timings. Dark and light red shading indicates the total $\sigma$ and 2$\sigma$ uncertainties. The dashed green lines give the 2$\sigma$ range for the intrinsic solar contribution (source scatter).}
\label{fig:DMs_preDiffPhases}
\end{figure}

We take the timing of 4 years and 8 years into cycles 23 and 24 as examples. Figure \ref{fig:DMs_preDiffPhases} shows the predicted time evolutions of the axial dipole moments for cycle 23 and cycle 24 at 4 years and 8 years into the cycles. Synoptic magnetograms CR1961 and CR2013 from MDI and CR2130 and CR2184 from HMI are used as the initial magnetic field, respectively. Solid green lines show the average of 50 SFT simulations with random sources. Dark and light red shading indicate the total $\sigma$ and 2$\sigma$ uncertainties, which correspond to the probabilities of 68\% and 95.4\% for the observed axial dipole moment to be within the error ranges, respectively. The dashed green lines give the 2$\sigma$ range for the intrinsic solar contribution (source scatter). The errors from MDI measurements are significant and are larger than that from HMI measurements. This also demonstrates the big effects of observed magnetograms on the prediction.

Figure \ref{fig:DMs_preDiffPhases} shows the following results. Firstly, the observed values are always within the 2$\sigma$ uncertainty, although statistically it is possible that the observation is out of the 2$\sigma$ range with the probability of 4.6\%. This demonstrates the effectiveness of prediction method. Secondly, the error ranges increase with time. The error ranges of the predicted axial dipole moment at the end of the cycles are smaller when the prediction is made later. This is due to the accumulative effect of the random source emergence. Thirdly, the absolute values of the predicted mean axial dipole moments decrease when the prediction was made at a later time. Cycle 23 had a deep minimum with very low axial dipole moment. \cite{Jiang2015} found that the weakness of the cycle 23 minimum are mainly caused by a number of bigger bipolar regions emerging at low latitudes with a ¡°wrong¡± (i.e., opposite to the majority for this cycle) orientation of their magnetic polarities in the north-south direction, which impaired the growth of the polar field. The abnormal emergences are supposed to be emerged during cycle 24 as well. Hence the mean value at cycle minimum decreases with time in both cycle 23 and cycle 24. The corresponding predictions of the following cycle strength decrease with time as well. But it is possible that the predicted mean values of the axial dipole moment increase with time if the peculiar BMR emergences (large areas and tilts at low latitudes) with positive contributions to the axial dipole moment surpass the peculiar BMR emergences with negative contributions to the axial dipole moment. In such case, a stronger subsequent cycle will be proceed.

\section{Correlation between the dipole moment at cycle minimum and the subsequent cycle strength}
\label{sec:DM_dataset}
The prediction of the subsequent cycle depends on the correlation between the dipole moment at cycle minimum and the following cycle strength. The physical base has been clarified in Section \ref{sec:introduction}. This correlation is a key ingredient to affect the prediction accuracy of the  subsequent cycle. A long-term homogeneous dipole moment dataset, which is directly calculated based on the observed synoptic magnetograms, is expected. The earliest available synoptic magnetograms are from Mount Wilson Observatory (MWO) from July 1974 until December 2012 \footnote{http://obs.astro.ucla.edu/intro.html}. Another long-term dataset which is widely used in the solar cycle prediction is the magnetic measurement from Wilcox Solar Observatory (WSO, continuously available from April 1976 until present \footnote{http://wso.stanford.edu/synopticl.html}). We use the axial dipole moment calculated based on MDI synoptic maps to cross-calibrate the other 3 axial dipole moments calculated based on MWO, WSO and HMI synoptic maps.

Figure \ref{fig:DMs_calibrations} shows the cross-calibrations of the axial dipole moment from MDI ($D_{\textrm{MDI}}$) with that from MWO ($D_{\textrm{MWO}}$, left panel), from WSO ($D_{\textrm{WSO}}$, middle panel) and from HMI ($D_{\textrm{HMI}}$, right panel). The linear fits give their relations $D_{\textrm{MDI}}=0.17+3.98D_{\textrm{MWO}}$, $D_{\textrm{MDI}}=0.14+5.42D_{\textrm{WSO}}$, and  $D_{\textrm{MDI}}=1.3D_{\textrm{HMI}}$. The correlation coefficients are 0.94, 0.96 and 0.89 respectively.

\begin{figure}[!htp]
 \centering
\includegraphics[scale=0.5]{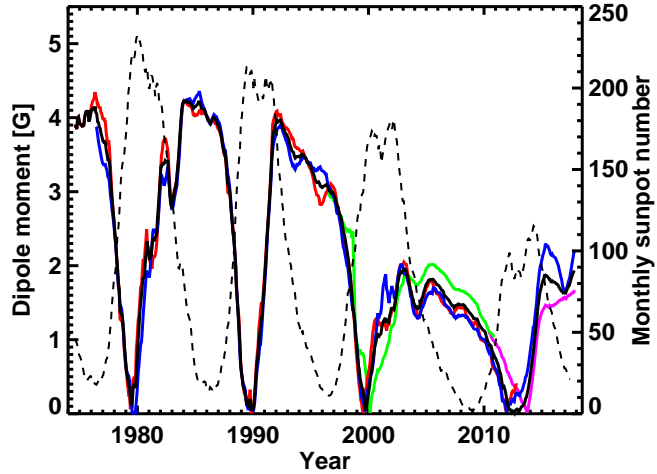}
\caption{Time evolutions of the absolute values of the 13-month running averaged axial dipole moments from the calibrated WSO (blue curve), MWO (red curve), HMI (pink curve) and MDI (green curve) and the monthly averaged sunspot number (black dashed curve). The homogeneous axial dipole moments calculated by average of the available values is shown in solid black curve.}
\label{fig:DMs_obs}
\end{figure}

\begin{figure}[!htp]
 \centering
\includegraphics[scale=0.5]{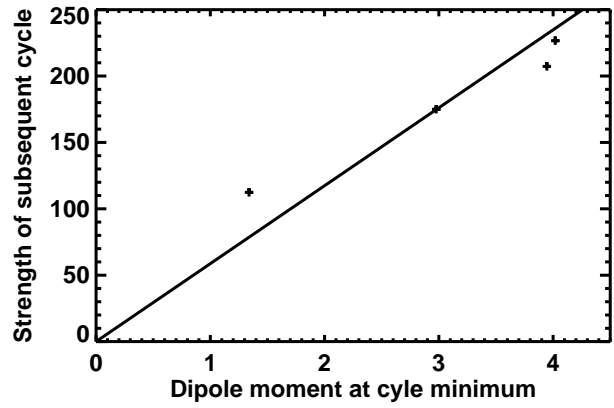}
\caption{Correlation between the axial dipole moment at cycle minimum $D_n$ and the subsequent cycle strength $S_{n+1}$. The correlation coefficient $r$ and the confidence level $p$ are $r=0.99$ and $p=0.045$, respectively.}
\label{fig:DM_SN_corr}
\end{figure}

Figure \ref{fig:DMs_obs} is the time evolution of the calibrated axial dipole moment after 13-month running averages in solid curves overplotted with sunspot number in dashed curve from 1974 onwards. We derive the time series of the averaged axial dipole moment $\overline{D}(t)$ in the following way. If there are more than one observations, we do the average over the different values. During the early time, we just take the values from MWO observations since only MWO observation is available. The 13-month running average of  $\overline{D}(t)$ is regarded as the homogeneous axial dipole moment dataset to derive the correlation between the dipole moment at cycle minimum $D_{\textrm{min}}^{n}$ and the following cycle strength $S_{n+1}$. To further reduce the random noise, we use the averaged value of 7 CRs around each minimum to get $D_{\textrm{min}}^{n}$. The cycle strength $S_{n+1}$ is the average of 7 months around each maximum of the smoothed sunspot data. The correlation between the homogenous dipole moment at cycle minima $D_{\textrm{min}}^{n}$ and the following cycle strength $S_{n+1}$ is shown in Figure \ref{fig:DM_SN_corr}, which indicates
\begin{equation}
\label{eq:DM_Sn+1}
S_{n+1}=58.7*D_{\textrm{min}}^{n}.
\end{equation}
Although there are only 4 points, the correlation coefficient is $r=0.99$ with the corresponding confidence level $p=0.045$. The sudden jump of the WSO observations during 2016-2017 in Figure \ref{fig:DMs_obs} is due to the reduced WSO polarization sensitivity \footnote{http://wso.stanford.edu/}. This time period has no effect on the correlation.

\section{Predictability of the subsequent cycle}
\label{sec:n+1 cycle predictability}
\subsection{Prediction of the subsequent cycle at different phases of a cycle}
We can predict the possible sunspot emergence of an ongoing cycle by random realizations of the features of sunspot emergence given in Section \ref{sec:CycleShape} a few years after the start of a cycle. With the SFT simulations and the observed synoptic magnetograms as the initial state of the magnetic field, the possible large-scale field evolution over surface can be obtained. Based on the correlation between the dipole moment at cycle minimum and the following cycle strength constrained in Section \ref{sec:DM_dataset} combined with the features of sunspot emergence given in Section \ref{sec:CycleShape}, the predictability of the shape of the subsequent cycle can be given.

Solar cycles also vary in length. Strong cycles tend to be shorter and vice versa, but with a large scatter \citep{Solanki2002, Du2006, Vaquero2008}. Some of this is due to the overlapping of sunspots from adjacent cycles. In this paper we ignore the cycle length variation and assume that all the cycles have an 11-year cycle period. If we know the maximum of a cycle $S_n$, we may derive the profile of the cycle based on the simplified formula of HRW94, which is just a function of $a$ relevant to cycle amplitude $S_n$. The parameter $t_0$ is not required. The correction function $\overline{\triangle r(t)}$ is still required for the improvement of the later phase. We only predict the systematic part of the smoothed sunspot number $\overline{F_{\textrm{sm}}}^S(t)$ for the prediction of the subsequent cycle, since the standard deviation of the prediction is usually larger than the intrinsic short term variability. The uncertainty results from the uncertainty of the axial dipole moment at the end of the cycle.

The formula to describe the subsequent cycle is
\begin{equation}\label{eq:r3}
\overline{F_{\textrm{sm}}}^S(t)=f_2(t)+\overline{\triangle r(t)}f_2(t),
\end{equation}
where
\begin{equation}\label{eq:fit3}
f_2(t)=\frac{a t^3}{\exp[t^2/b^2]-c},
\end{equation}
and $\overline{\triangle r(t)}$ corresponds to the formula of Eq.(\ref{eq:r1a}).
The fits of cycles 12-24 using Eq.(\ref{eq:fit1}) indicate that the relationship between $S_n$ and $a$ is
\begin{equation}\label{eq:Sn_a}
S_n=9072.8a^{0.706}.
\end{equation}
The parameters $b$ and $c$ are the same as that in Eq.(\ref{eq:cyFit}). The deviation of the maximum sunspot number given by Eq.(\ref{eq:fit3}) from $S_n$ is within 5.

\begin{figure}[!htp]
 \centering
\includegraphics[scale=0.4]{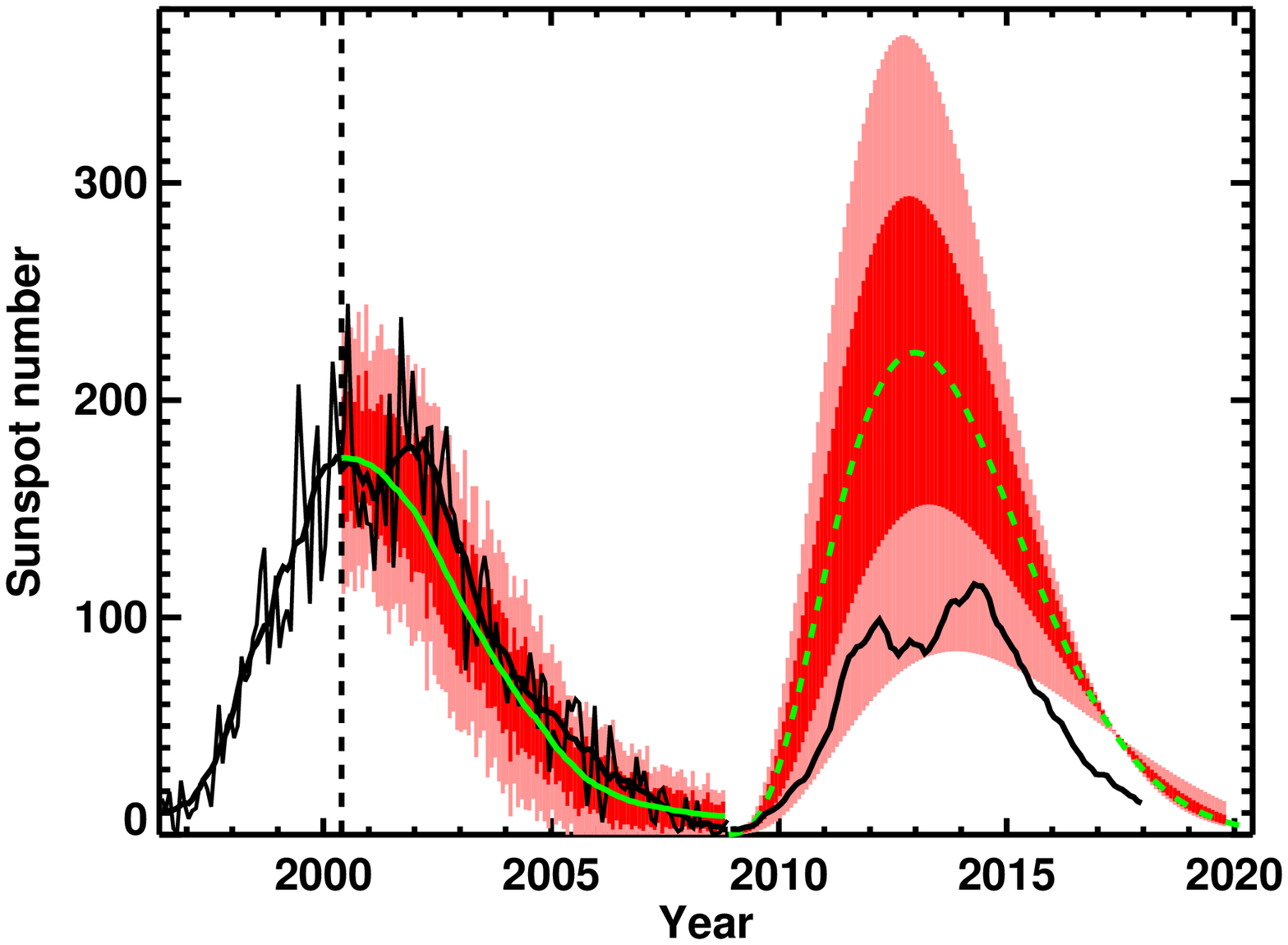}
\includegraphics[scale=0.4]{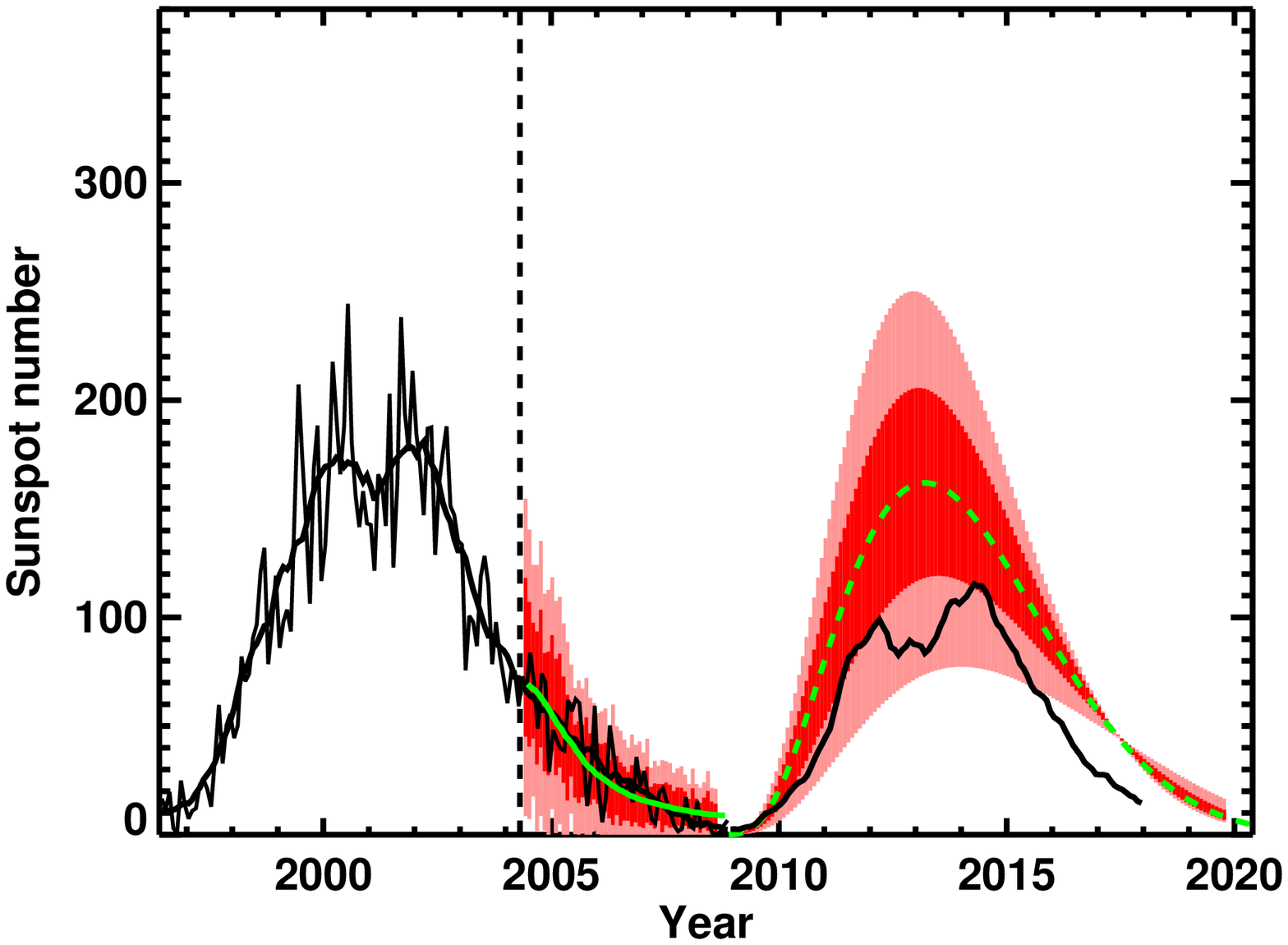}
\includegraphics[scale=0.4]{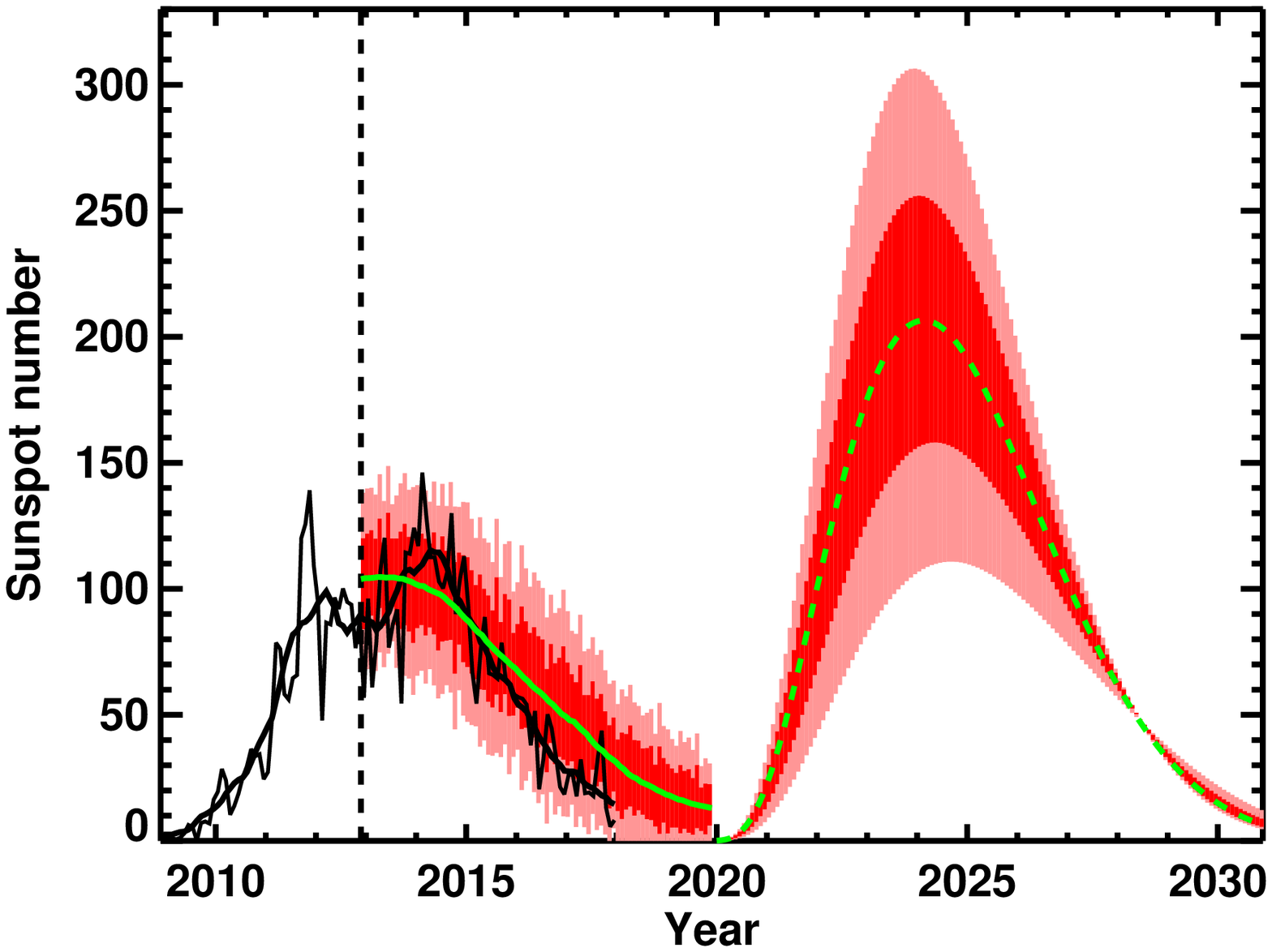}
\includegraphics[scale=0.4]{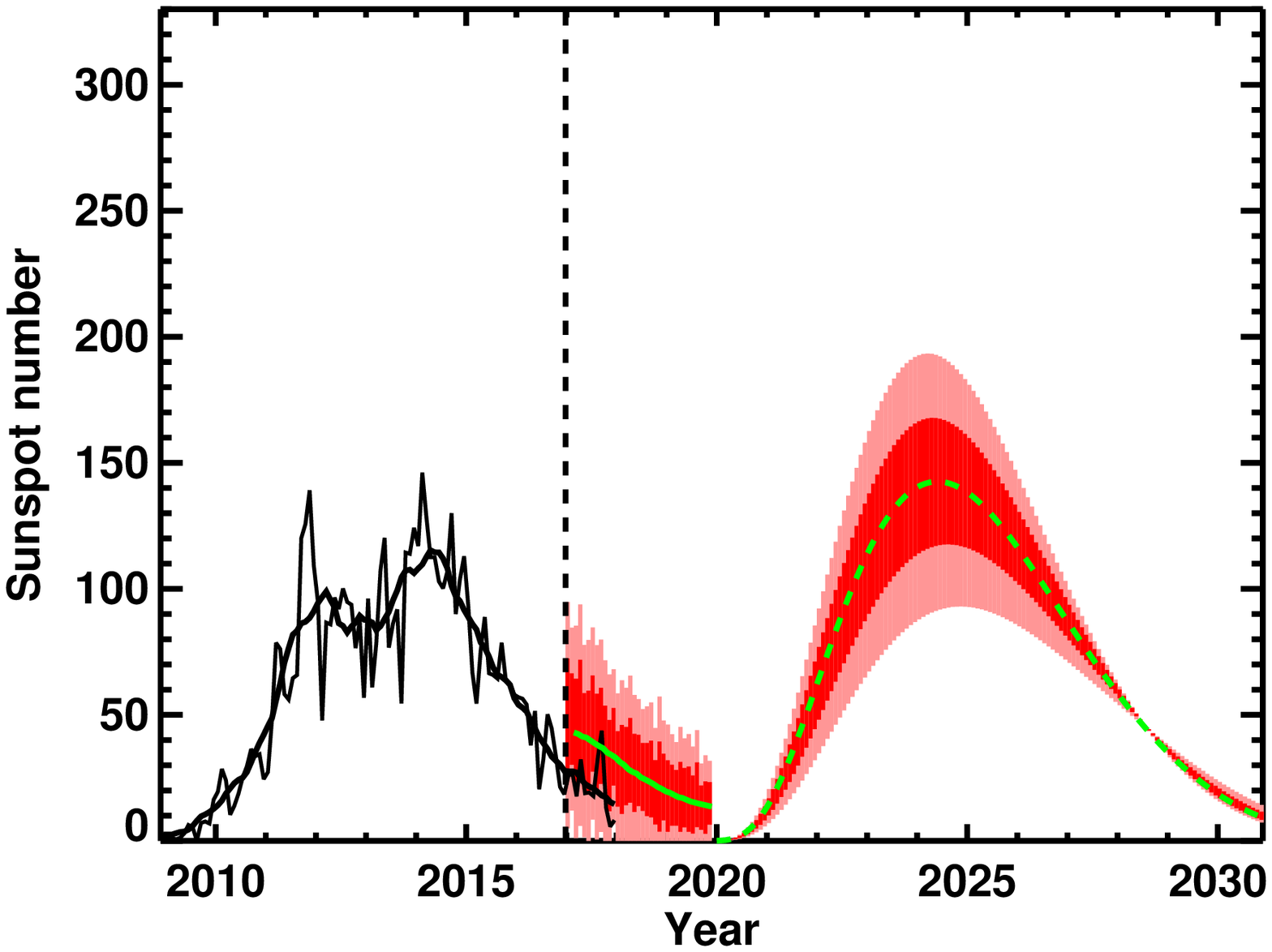}
\caption{Examples of the predictions of the subsequent cycle at different times after the start of the cycle. Left panels: 4 years into cycle 23 (upper one) and cycle24 (lower one); Right panel: 8 years into cycle 23 (upper one) and cycle 24 (lower one). The green curves show the averages of the 50 random realizations of the sunspot emergence. Dark and light red shading indicates the total $\sigma$ and 2$\sigma$ uncertainty. For ongoing cycles, the $\sigma$ range denotes the uncertainty of the monthly sunspot number. And for subsequent cycles, the $\sigma$ range denotes the uncertainty of the smoothed sunspot number.}
\label{fig:SN2cy_preDiffPhases}
\end{figure}

Figure \ref{fig:SN2cy_preDiffPhases} shows some examples illustrating the predictability of the subsequent cycle, starting from different phases of a cycle. The timings for the predictions are the same as the ones in Figure \ref{fig:DMs_preDiffPhases}. From Figure \ref{fig:DMs_preDiffPhases}, we can derive $\overline{D}_{n}+2\sigma^D_{n}$, $\overline{D}_{n}+\sigma^D_{n}$, $\overline{D}_{n}$, $\overline{D}_{n}-\sigma^D_{n}$, and $\overline{D}_{n}-2\sigma^D_{n}$ at the cycle $n$ minimum. We derive the corresponding amplitude of the subsequent cycle strength, $\overline{S}_{n+1}+2\sigma^S_{n+1}$, $\overline{S}_{n+1}+\sigma^S_{n+1}$, $\overline{S}_{n+1}$, $\overline{S}_{n+1}-\sigma^S_{n+1}$, and $\overline{S}_{n+1}-2\sigma^S_{n+1}$ using Eq.(\ref{eq:DM_Sn+1}). Still the $\sigma^S_{n+1}$ and 2$\sigma^S_{n+1}$ ranges correspond to the probabilities of 68\% and 95.4\% for the actual sunspot number to be within the error ranges. Based on Eqs.(\ref{eq:r3}), (\ref{eq:fit3}), and (\ref{eq:Sn_a}), the profiles of the smoothed sunspot number in the subsequent cycle $n+1$ can be obtained. We see that the observed sunspot number in cycle 24 is within $\sigma$ to $2\sigma$ range of the predicted result whatever the prediction is at the early time, i.e., the 4th year or the later time, i.e., the 8th year. The error range decreases with time. The deviation between the observation and the prediction during the decline phase is expected since the prediction given here is only the systematic part. As shown in Table \ref{tab:GoodnessOfFit}, the systematic part sometimes shows a deviation from the observed sunspot number evolution (but the observations still are within the given error range). The updated prediction of the cycle based on a later magnetogram and the random flux source is presented in the following subsection.

\subsection{Updated predictability of cycle 25}
\begin{figure}[!htp]
 \centering
\includegraphics[scale=0.4]{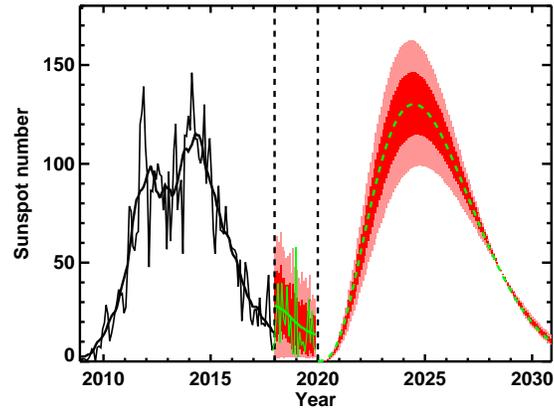}
\caption{Predictability of cycle 25 using the synoptic magnetogram CR2198 (the last CR of 2017). The curves are the same as in Figure \ref{fig:SN2cy_preDiffPhases}. The amplitudes of cycle 25 based on the smooth sunspot number is $125\pm32$.}
\label{fig:LatestPrdc2}
\end{figure}

\begin{figure}[!htp]
 \centering
\includegraphics[scale=0.4]{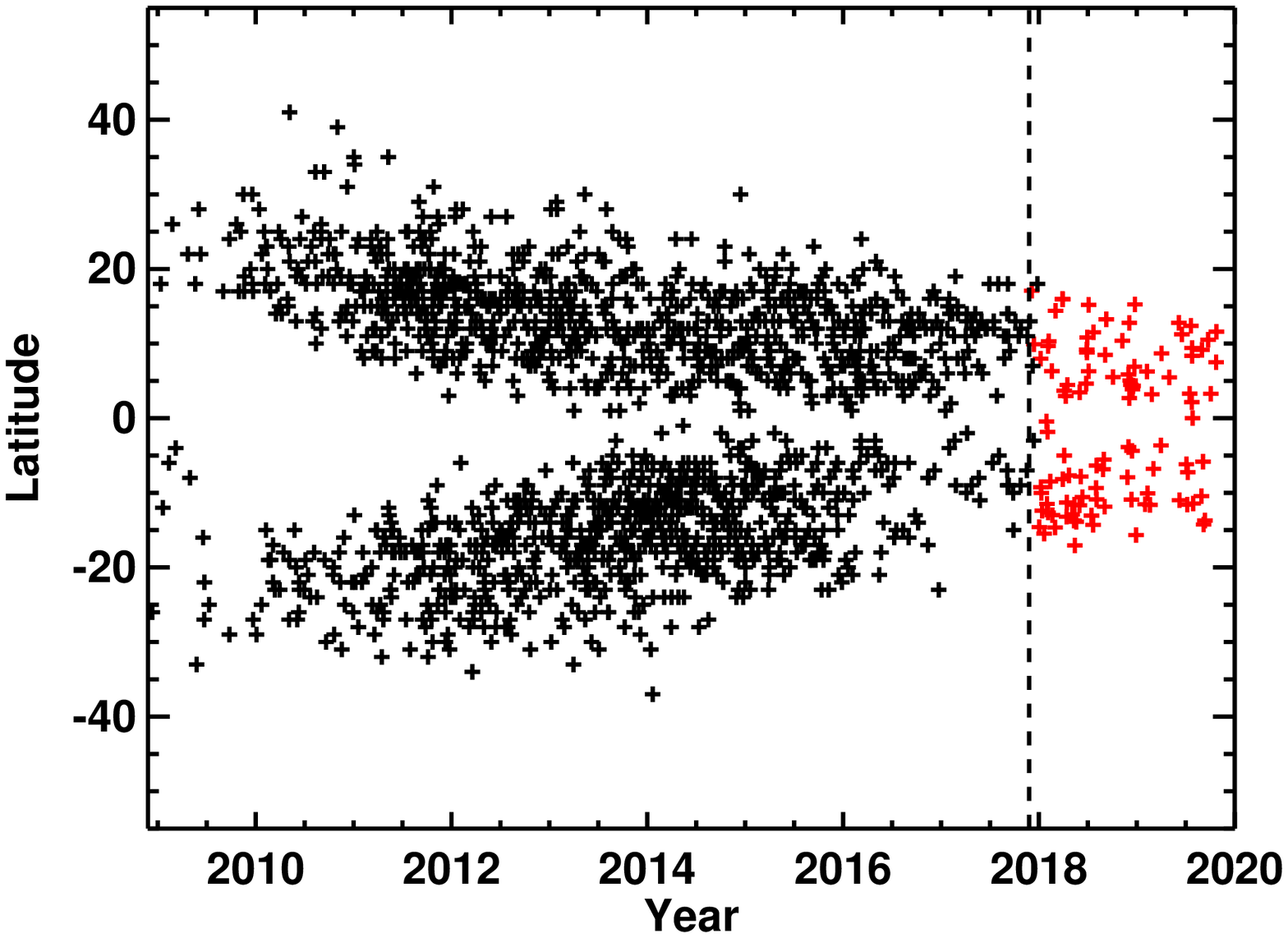}
\includegraphics[scale=0.4]{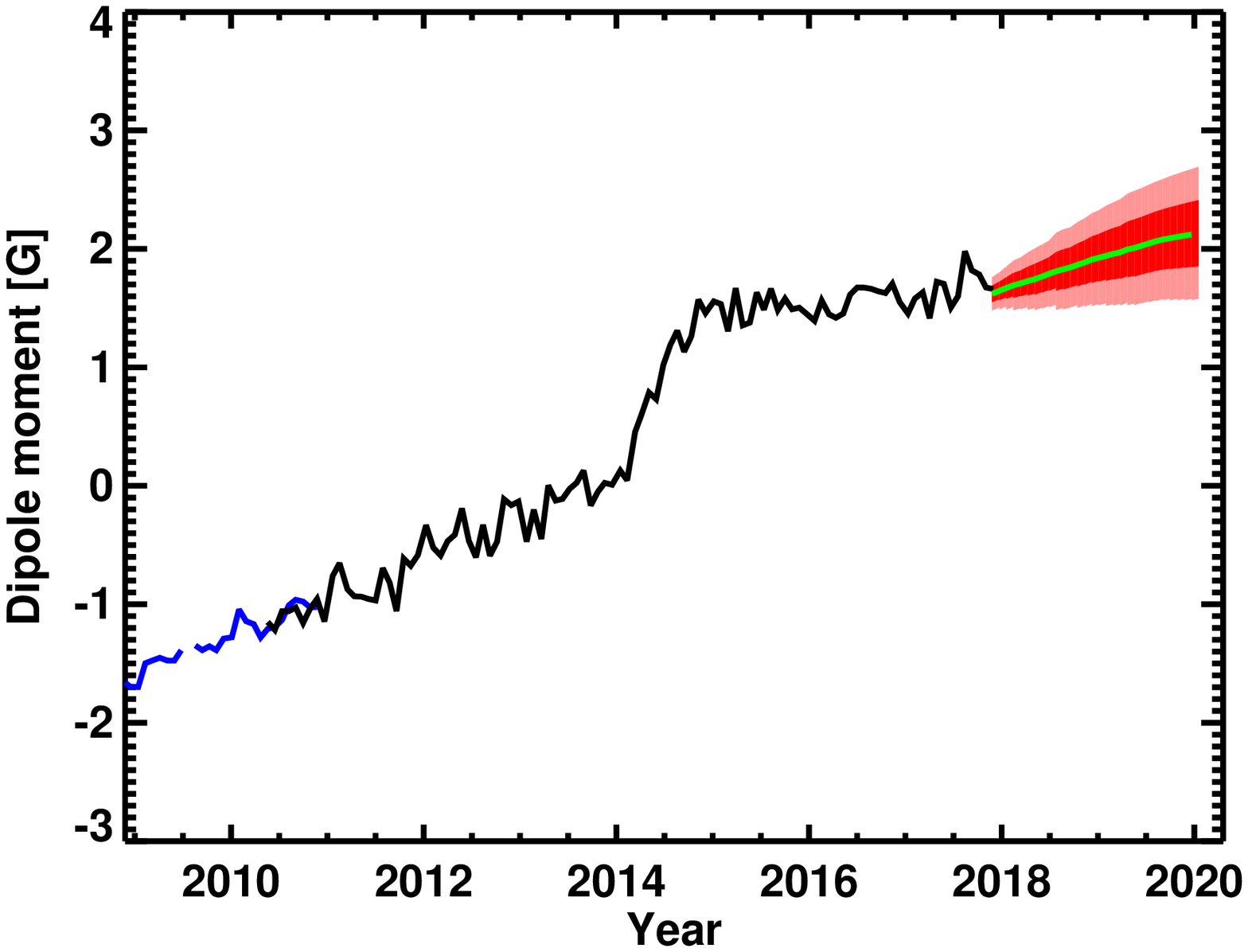}
\includegraphics[scale=0.4]{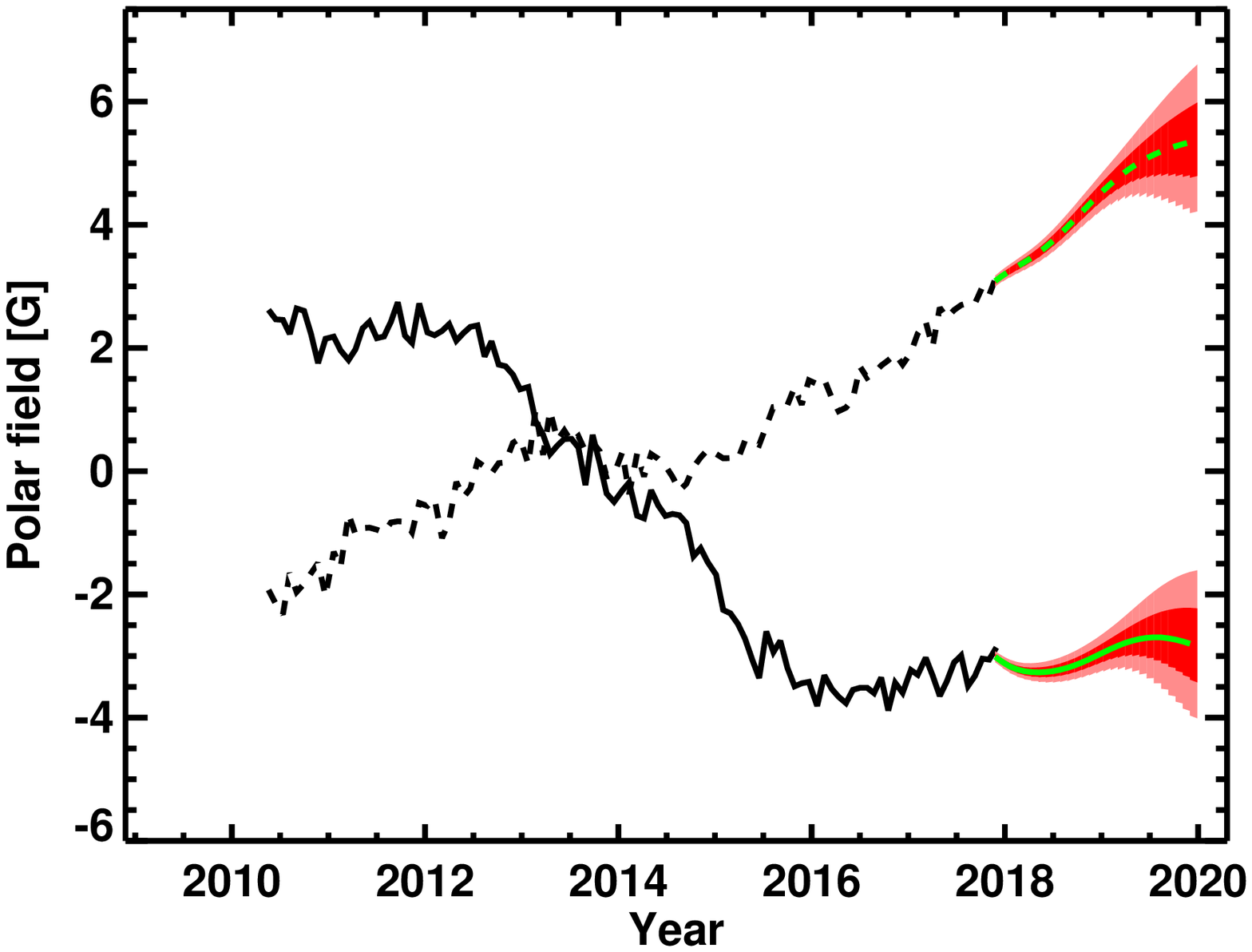}
\includegraphics[scale=0.4]{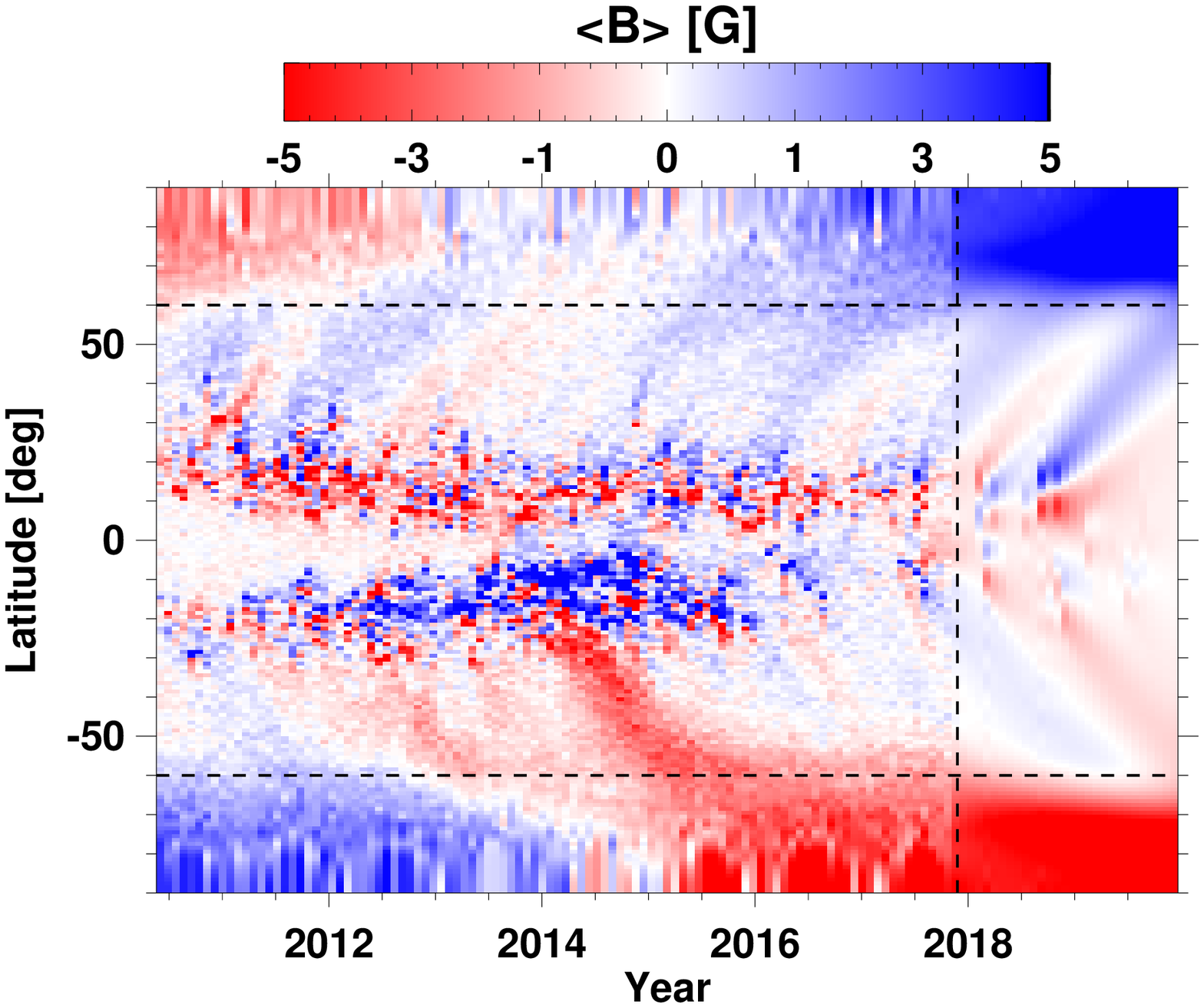}
\caption{Prediction of the large-scale field evolution over the solar surface during the rest of the ongoing cycle 24 using the synoptic magnetogram CR2198 (December 2nd -- December 31st, 2017) as the initial field. Upper left: the butterfly diagram. The observed sunspot groups are in black and the predicted spots by one random realization are in red. Upper right: Axial dipole field evolution; Lower left: polar field evolution, north polar field in dashed curve and south polar field in solid curve. In upper right and lower left panels, the green curves are the averaged values of 50 random realizations. The light and dark red shade regions correspond to $\sigma$ and 2$\sigma$ uncertainty range; Lower right: longitudinal averaged surface field evolution combined with the HMI observations (before the vertical line) and the simulation (after the vertical line) corresponding to the random realization in Upper left panel.}
\label{fig:LatestPrdc}
\end{figure}

Here we take the synoptic magnetogram, CR2198 (December 2nd -- December 31st, 2017), as the initial condition to investigate the large-scale field evolution over the solar surface during the rest of cycle 24 and the possible profiles of cycle 25. A possible (one from our Monte-Carlo ensemble) monthly sunspot emergence pattern during the rest of the ongoing cycle 24 is shown in Figure \ref{fig:LatestPrdc2}. The systematic part of the smoothed sunspot number is shown in the solid green curve. It is higher than the current observation. But the deviations are within 2$\sigma$ range of the fluctuation. The thin green curve is one realization of the smoothed sunspot number. The corresponding time evolution of the latitudinal location of the sunspot groups (observed in black and predicted in red) is presented in the upper left panel of Figure \ref{fig:LatestPrdc}. The detailed description of Figure \ref{fig:LatestPrdc} which is the predicted large-scale field evolution during the rest of cycle 24 is as follows.

The upper right panel of Figure \ref{fig:LatestPrdc} shows the time evolution of the axial dipole moment from MDI and HMI observations in blue and in black before the end of 2017 and the predicted one after 2017. The mean value will increase from 1.66 G at 2018.0 to 2.12G with 2$\sigma$ range of 0.55G at 2020. We also calculated the averaged polar field over $\pm60^{\circ}$ to $\pm75^{\circ}$ latitudes to understand the different evolutions of the two hemispheres. The northern polar field is in dashed curve and the southern polar field is in solid curve. The green curves show the expected values and the dark and light red shades show the 1$\sigma$ and 2$\sigma$ range of the predicted polar field. During the first one year, the error range is very small. The polar field is determined by the initial conditions. The northern and the southern polar fields are almost in balance with values of 3.17G and -2.86G. The southern polar field will keep almost flat with the values varying to -2.94G, and the northern polar field will have a large increase to 4.56G at the end of 2018. The averaged northern polar fields keep increasing and the averaged southern polar fields keep stable until the end of the cycle. The mean values by then are -2.89G and 5.40G with 2$\sigma$ range of 1.23G, respectively. The reason for the increasing northern polar field and the stable southern polar field can be explained by the lower right panel of Figure \ref{fig:LatestPrdc}, which is the longitudinal averaged surface field evolution combined with the HMI observations (before the vertical line) and the simulation (after the vertical line) corresponding to the random realization in upper left panel. There are strong poleward positive plumes starting from second half of 2017. The positive plume in the northern hemisphere further increased the positive polar field. The positive plume in the southern hemisphere prevented the increase of southern negative polar field. Hence it keeps almost stable. The two notable plumes mainly result from two big ARs, i.e., AR12674 and AR12673 \citep{Yang2017, Yan2018}. They occurred on the solar disc with large tilts around September 5th, 2017 locating at northern and southern hemisphere respectively. According to \cite{Jiang2014b, Jiang2015}, the bigger bipolar regions emerging at low latitudes have significant effects on the large-scale field evolution. Their effects on the solar cycle will be studied in detail in a separated study.

The possible evolution of the smoothed sunspot number based on the axial dipole moment at the end of cycle 24 is given in Figure \ref{fig:LatestPrdc2}. The expected maximum amplitude of cycle 25 is 125, which is about 10\% higher than current cycle 24. The 2$\sigma$ range is 32, which means that the possibility of the amplitude of cycle 25 above 93 is 95.4\%. Cycle 25 most probably is a normal cycle, rather than the Maunder minimum period. The possible profiles of cycle 25 denoted by the shaded region obey the empirical relations given in Section \ref{sec:CycleShape}.

In order to avoid the effects due to the unexpected problems from synoptic magnetograms, we also repeated the predictions based on synoptic magnetograms, CR2196 and CR2197. The results are similar.

\section{Summary and discussion}
In this paper, we have developed a scheme to investigate the predictability of the solar cycle over one cycle. The scheme includes three steps. Firstly, empirical properties of the solar cycle are used to predict the possible sunspot emergence for an ongoing cycle. Then the SFT model is adopted to predict the possible large-scale field evolution over the surface, including the polar field at the end of the cycle. Finally, the correlation between the polar field and the subsequent cycle strength and empirical properties of the sunspot emergence are applied to get the possible profiles of the subsequent cycle. The scheme is verified by past cycles and is applied to predict the possible profiles of cycle 25. The results show the cycle 25 strength of $125\pm32$ (2$\sigma$ uncertainty range), which is about 10\% stronger than cycle 24 based on the mean value.

Comparing with the existing methods of the solar cycle prediction, the main progress of the current scheme is as follows. Firstly, the prediction scope is extended over one cycle. Secondly, the profiles of solar cycles during the whole prediction time period, rather than only the cycle amplitude, can be given. Thirdly, the uncertainty due to the randomness of the sunspot emergence and the data assimilated to the model during the whole prediction time period is given. Fourthly, not only the monthly and smoothed sunspot number but also the emergence properties, including the size, location, and tilt of sunspot groups, can be given. Fifthly, although the BL dynamo model is not directly used in the model, the empirical properties we used have the solid dynamo origin. These five properties distinguish the scheme from all the current existing attempts of the solar cycle predictions.

For one source of prediction uncertainties, the imperfect measurement of the initial magnetogram, we only estimate the error from the net flux density. The error due to other possible problems, like the center-to-limb correction of the magnetic saturation is not dealt with in this paper. Potential problems of the initial magnetograms can degrade the predictive ability. We have explicitly assumed that the scatter in the tilt angle is random in nature, consistent with the idea that it is due to turbulent buffeting by the convective motions. If there is a significant deterministic chaotic component in the dynamo process, which we currently do not see evidences for, the model can be improved. Conversely if there is some chaotic or random process operating on long timescales and which exceeds the randomness introduced by the scatter in the tilt angles, then our prediction can fail. Furthermore, we assumed all cycles have 11-year cycle period, and ignored the cycle overlap. So there remains room for improving the scheme, although these improvements are likely to be small compared to the uncertainties due to the tilt-angle scatter which we explicitly dealt with in this paper.

\begin{acknowledgements}
We are grateful to Robert Cameron, Sami Solanki, and Manfred Sch\"{u}ssler for stimulating discussions. Robert Cameron also helped to improve the language. SOHO is a project of international cooperation between ESA and NASA. The SDO/HMI data are courtesy of NASA and the SDO/HMI team. The sunspot records are courtesy of WDC-SILSO, Royal Observatory of Belgium, Brussels. The National Solar Observatory (NSO)/Kitt Peak data used here are obtained cooperatively by NSF-NOAO, NASA/Goddard Space Flight Center, and the NOAA Space Environment Laboratory. NSO/SOLIS data were courtesy of NISP/NSO/AURA/NSF. We acknowledges the support by the National Science Foundation of China (grants 11522325, 11573038, and 41431071) and by the Fundamental Research Funds for the Central Universities of China.
\end{acknowledgements}


\newpage

\end{document}